%% file: main.tex
\documentclass{aa}  

\usepackage{graphicx}
\usepackage{txfonts}
\usepackage{lipsum}
\usepackage{placeins}

\usepackage{amsmath}
\allowdisplaybreaks
\usepackage{bbold}

\usepackage{subcaption}
\input{commands.tex}

\usepackage{xparse}
\input{events.tex}

\ExplSyntaxOn
\NewDocumentCommand{\fullevent}{m}
 {
  \prop_item:Nn \g_aliases_prop {#1} 
 }
\ExplSyntaxOff

\usepackage{tikz}
\newcommand{\cir}[2]{\begin{tikzpicture}
\node[draw,scale=0.7,circle,{#1},{#2}] at (0,0) {}; 
\end{tikzpicture}}
\definecolor{GW230814}{HTML}{af3800}
\definecolor{GW231123}{HTML}{00916e}
\definecolor{GW231224}{HTML}{f3ca40}
\definecolor{GW231226}{HTML}{054a91}
\definecolor{GW250114}{HTML}{533745}

\usepackage{hyperref}

\definecolor{referee}{HTML}{d70040}
\newcommand{\revised}[1]{#1}

\definecolor{comment}{HTML}{00916e}

\begin{document}

\title{Remnant recoil and host environments of \GWTCFOUR binary black-hole mergers}

\author{J.~Llobera-Querol\inst{1}\fnmsep\thanks{Corresponding author: joan.llobera@uib.cat}
    \and E.~Hamilton\inst{1}
    \and N.~Singh\inst{1}
    \and M.~Colleoni\inst{1}
    \and F.~A.~Ramis~Vidal\inst{1}
    \and A.~Askar\inst{2}
    \and T.~Bulik\inst{3}
    \and A.~Olejak\inst{4}
    \and S.~Husa\inst{5}\fnmsep\inst{1}
    \and Y.~Xu\inst{1}
    \and J.~Valencia\inst{1}
}

\institute{\UIB
    \and \WarsawCopernicus
    \and \WarsawObservatory
    \and \MPIAstro
    \and \ICE}

\date{Received \today}

\abstract
    {Determining the astrophysical origin of binary black holes and whether merger remnants are retained in their birth environments is essential for understanding hierarchical mergers and the growth of intermediate-mass black holes.}
    {We identified the gravitational-wave (GW) events most consistent with \revised{dense-cluster origin} and assessed whether their merger remnants are retained in globular clusters, nuclear star clusters, or galactic potentials.}
    {We considered the 84 events consistent with binary-black-hole (BBH) mergers from the first part of the fourth observing run (\OFOURa) of the LIGO-Virgo-KAGRA (LVK) GW detector network, and 3 selected events from the second part (\OFOURb).
    We compared parameter-estimation posteriors with synthetic population models for field and cluster binaries using Bayes factors, accounting for the relative abundances of these formation channels in the local Universe.
    We computed recoil-velocity posteriors for all events using the \IMRPhenomXPNR waveform model, which incorporates multipole asymmetries.}
    {We identified five events \revised{whose inferred intrinsic parameters show preference for the adopted dense-cluster population models over the considered field-binary populations}, including the most massive \OFOURa event \fullevent{GW231123}, while \revised{finding no robust preference for a dense-cluster origin for} the high-$\chieff$ \OFOURb event \fullevent{GW241011} \revised{within the adopted population models}.
    Typical recoil velocities of analyzed events are of order a few hundred $\kmstight$, with extended high-velocity tails.
    These kicks suggest that merger remnants are likely ejected from typical globular clusters, while retention in nuclear star clusters remains possible but not guaranteed.}
    {\revised{Within the adopted models and assumptions, our results suggest that efficient} hierarchical growth \revised{may be challenging in typical} globular clusters, whereas nuclear star clusters remain viable environments for repeated mergers.
    Although results depend on the adopted astrophysical population models, this analysis highlights the importance of improved and larger population models, as well as higher-quality detections enabled by future developments in GW detectors.}


\maketitle
\nolinenumbers

\section{Introduction}

Understanding the formation channels of compact objects is one of the prime scientific objectives of GW astronomy. 
Here we focus on BBHs,
where the latest catalog of transient signals, \GWTCFOUR \citep{LIGOScientific:2025slb},
has more than doubled the number of observed signals \citep[see][for an introduction and description of data analysis methods]{LIGOScientific:2025hdt,LIGOScientific:2025yae}.
This has allowed for the confirmation of features in the intrinsic parameters that suggest contributions from multiple formation channels and subpopulations 
\citep{LIGOScientific:2025pvj}.
Indeed, it is widely accepted that BBHs may form through a variety of astrophysical channels, including in particular the ``classical'' evolution of isolated binary stars, but also dynamical encounters in dense environments, such as globular clusters (GCs) and nuclear star clusters (NSCs). 
For recent reviews see e.g.~\citet{Mapelli:2020vfa,Mapelli:2021taw}.

While many detected events are broadly consistent with isolated binary evolution in low-density environments, formation in dense stellar systems is expected to contribute a non-negligible fraction of the observable BBH population
\citep{LIGOScientific:2025pvj,Tong:2025xir,Plunkett:2026pxt,Farah:2026jlc,Vijaykumar:2026zjy,Gerosa:2021mno,Banagiri:2025dmy,Sedda:2026xqr}.                
Moreover, several events have been proposed as possible candidates for hierarchical mergers,
including \fullevent{GW231123} in \OFOURa \citep{Li:2025fnf,Li:2025pyo,Paiella:2025qld,Passenger:2025acb,Liu:2025yok};
\fullevent{GW241011} and \fullevent{GW241110} in \OFOURb \citep{LIGOScientific:2025brd,Li:2025odl};
and \event{GW190412} \citep{Hamers:2020huo,Gerosa:2020bjb},
\event{GW190521} \citep{Romero-Shaw:2020thy,Sedda:2021abh,LIGOScientific:2020ufj,Anagnostou:2020tta,Mahapatra:2024qsy,Alvarez:2024dpd},
and \event{GW190814} \citep{Lu:2020gfh}
in \OTHREE \citep{Liu:2020gif}.
Such candidates are typically identified through distinctive signatures of second-generation binaries, such as large component masses or high spin magnitudes.
However, these features are not uniquely associated with hierarchical formation: high spins can also arise in isolated binaries \citep{Belczynski:2017gds,Bavera:2020inc,Olejak:2021iux},
and first-generation binaries formed in clusters may display other characteristics that need to be identified for a proper environment distinction.
Hierarchical mergers are expected to represent only $\sim0.2\%-2\%$ of all BBH mergers in GCs, with rates roughly an order of magnitude larger in NSCs, \citep[see e.g.][]{Li:2022gly,Mapelli:2020xeq}.

In this work, we identified events for which a cluster origin \revised{is preferred},
by combining state-of-the-art catalogs of BBHs formed in both dense stellar environments and isolated binaries, with parameter-estimation (PE) posteriors from GW observations.
We used the predicted distributions of intrinsic parameters of merging compact binaries from these catalogs to assess the consistency of each BBH detected by the LVK Collaboration with different formation channels.

Our ability to distinguish formation channels remains limited by the relatively small number of detected events spread over a large parameter space, the typically broad posterior distributions on source parameters resulting from low signal-to-noise ratios, and further complicated due to degeneracies in inferred source parameters.
Consequently, our analysis focused on a subset of relatively well-measured intrinsic parameters: the total mass $M$, the mass ratio $q$, and the inspiral effective spin $\chieff$, since these are the parameters that appear in the leading terms in Post-Newtonian (PN) expressions valid in the inspiral for the gravitational waveforms \citep{Baird:2012cu} and $\chieff$ is conserved at 2PN order \citep{Racine:2008qv}.
Our analysis avoided parameters that, while potentially informative, are currently poorly constrained at present observational sensitivities, such as individual spin magnitudes~$\chi_i$, the precessing effective spin~$\chip$ or orbital eccentricity~$e$.

Compact objects and binaries in dense stellar environments may escape their host gravitational potential if their velocity exceeds the escape velocity.
BBHs may acquire substantial velocities through dynamical encounters, leading a significant fraction of systems to be ejected prior to merger
\citep{PortegiesZwart:1999nm, Askar:2016jwt,2023arXiv231112118A}.
Binaries that merge within the cluster can produce remnants that receive recoil kicks large enough to exceed the escape velocity, ejecting them from their host environments and thereby preventing subsequent hierarchical mergers \citep{Campanelli:2007ew,Campanelli:2007cga}.
Quantifying the fraction of merger remnants retained in clusters therefore has important implications, including hierarchical merger rates and the resulting BBH mass distributions.

Recoil velocities have previously been inferred directly from GW observations for \OTHREE events \citep{Varma:2020nbm}.
Individual systems have been identified as potential high-kick candidates, most notably \event{GW200129} with recoil velocities of $\sim 1500 \kms$ \citep{Varma:2022pld}.
Using the surrogate model \NRSurRemnant \citep{Varma:2018aht,Varma:2019csw}, \citet{Doctor:2021qfn} inferred recoil distributions and estimated that approximately $3\%$ ($46\%$) of \OTHREE merger remnants are retained in GCs (NSCs).
Similarly, \citet{Mahapatra:2021hme}, using fitting formulas for recoil velocities, estimated retention fractions of $\sim1-12\%$ and $\sim14-70\%$ for escape velocities of $50$ and $250\kms$, respectively.
Neither of these studies tries to identify which of the \OTHREE events come from clusters.
This background motivates extending recoil and retention analyses to the \GWTCFOUR population using state-of-the-art waveform models.

The \GWTCFOUR catalog \citep{LIGOScientific:2025hdt, LIGOScientific:2025yae, LIGOScientific:2025slb}, released by the LIGO-Virgo-KAGRA (LVK) Collaboration \citep{LIGOScientific:2014pky,VIRGO:2014yos,KAGRA:2020tym}, includes 84 new candidate BBH mergers detected during the first part of the fourth observing run (\OFOURa).
With the corresponding publicly available open data \citep{LIGOScientific:2025snk},
these events have been reanalyzed using phenomenological waveform models \citep{Xu:2025ajj}, including parameter inference with \IMRPhenomXPNR  \citep{Hamilton:2025xru}.
Here, we also include three additional \OFOURb events, for which public data are available: \fullevent{GW241011} and \fullevent{GW241110} \citep{LIGOScientific:2025brd}, and \fullevent{GW250114} \citep{LIGOScientific:2025rid}.

Here we identified the BBH mergers that are more likely to originate in dense stellar environments and then assessed the retention of their merger remnants in their host environments, potentially leading to hierarchical mergers. This analysis consists of two steps:
First, we compared PE posteriors for GW events obtained with \IMRPhenomXPNR to population catalogs for different formation environments to identify candidate cluster-origin events, providing a detailed analysis and discussing individual systems of interest.
Second, we computed recoil velocity distributions from posterior samples
and the retention probability of merger remnants in their host environments, including globular clusters, nuclear star clusters and typical galactic potentials.

The paper is organized as follows.
In Sect.~\ref{sec:2} we describe the waveform modeling, recoil computation and PE framework used in this work.
Section~\ref{sec:3} summarizes the population catalogs for dense stellar environments and field binaries used in our analysis.
In Sect.~\ref{sec:4}, we discuss which events \revised{show larger overlap with} dense environments.
Section~\ref{sec:5} presents recoil distributions for all events and retention probabilities for the selected systems.

For BBH component masses $m_1,m_2$ with $m_1\geq m_2$, we define the total mass $M=m_1+m_2$, and the mass ratio $q=m_2/m_1 \leq 1$.
Masses observed in the detector frame are redshifted with respect to source masses, following $m_{det} = (1+z) m_{source}$. Here we show source masses, while $q$ and $\chieff$ are not affected by this distinction. 
The dimensionless spin vectors $\chii$ have magnitudes $\chi_i=|\chii| \leq 1$ and their components parallel and perpendicular to the orbital angular momentum of the binary $\Lhat$ are $\chiipar = \chii \cdot \Lhat$ and $\chiiperp = \| \chii \times \Lhat \|$.
Throughout this work, we used the Planck18 cosmology \citep{Planck:2018vyg} included in \astropy \citep{2022ApJ...935..167A}.

\section{Methods} \label{sec:2}

This section describes the theoretical framework on waveform modeling and PE required in this study.
We discuss waveform asymmetries and their implementation in \IMRPhenomXPNR, with the focus on accurate recoil velocities,
outline the computation of such remnant kicks,
and summarize the PE setup used to obtain posterior samples for the analyzed events. 

\subsection{Waveform modeling and multipole asymmetries}

The emission of gravitational radiation from a BBH system is fully characterized by the multipolar decomposition of the complex strain,
\begin{equation}
    h(t,\theta,\varphi) = \sum_{\ell,m} h_{\ell,m}(t) \ {}_{-2}\mathcal{Y}_{\ell,m}(\theta,\varphi)
,\end{equation}
where ${}_{-2}\mathcal{Y}_{\ell,m}(\theta,\varphi)$ is a basis of spin-weight $-2$ spherical harmonics, $\theta,\varphi$ are the polar and azimuthal angles of the binary in the sky,
and $h_{\ell,m}$ are the waveform modes.

For non-precessing binaries with spins aligned or anti-aligned with the orbital angular momentum $\Lhat$, the system is invariant under reflection across the orbital plane.
In a frame where the $z$-axis is aligned with $\Lhat$ this equatorial symmetry enforces the relation
\begin{equation} \label{eq:symmetry}
    h_{\ell,m}(t) = (-1)^\ell\ h^\ast_{\ell,-m}(t)
,\end{equation}
where $^\ast$ denotes complex conjugate.
This symmetry enforces exact cancellations in the linear-momentum flux perpendicular to the orbital plane.

When the component spins are misaligned with $\Lhat$, relativistic spin-orbit coupling induces precession and the equatorial symmetry is generically broken \citep{Apostolatos:1994mx}.
As a result, Eq.~\eqref{eq:symmetry} no longer holds, and the waveform develops intrinsic multipole asymmetries between $m$ and $-m$ modes.
These asymmetries are most pronounced during the late inspiral and merger, where precession effects and higher-order mode couplings are strongest, and are essential for generating substantial out-of-plane momentum flux and therefore large recoil velocities \citep{Bruegmann:2007bri}.
For more details on the effect on the waveform itself, see e.g. \citet{Arun:2008kb, Boyle:2014ioa}.

Precessing waveform models are commonly constructed in a co-precessing frame \citep{Schmidt:2010it,OShaughnessy:2011pmr,Boyle:2011gg}, in which the $z$-axis approximately follows the Newtonian orbital angular momentum of the binary. Therefore during the inspiral phase it is approximately described as a non-precessing system whose orientation evolves in time.
While this approach efficiently captures the leading effects of precession on the waveform morphology, if the underlying co-precessing waveform satisfies Eq.~\eqref{eq:symmetry}, the inertial-frame waveform will be missing the asymmetry and consequently will not accurately reproduce the gravitational emission from precessing systems.
Explicit asymmetry prescriptions must therefore be incorporated at the level of the co-precessing waveform model.

A convenient way to characterize equatorial symmetry breaking is through the combinations
\begin{equation}
    h^\pm_{\ell,m}(t) \equiv \dfrac{h_{\ell,m}(t) \pm (-1)^\ell h^\ast_{\ell,-m}(t)}{2}
,\end{equation}
where $h^+_{\ell,m}$ and $h^-_{\ell,m}$ represent the symmetric and antisymmetric contributions, respectively.
In aligned-spin systems, the antisymmetric components vanish identically, whereas in precessing systems they encode the degree of asymmetry in the system \citep{Mielke:2026xcm}.

Different waveform families implement asymmetries in distinct ways.
In frequency-domain phenomenological models, such as \IMRPhenomXOFOURa \citep{Thompson:2023ase} and \IMRPhenomXPNR \citep{Hamilton:2025xru}, asymmetries are introduced in the dominant $(2,\pm2)$ multipoles through prescriptions calibrated to numerical relativity (NR).
The asymmetric amplitude is modeled as a ratio to the symmetric amplitude, while analytic relations between the symmetric and antisymmetric phases are used during inspiral and ringdown, following \citet{Ghosh:2023mhc}.
In the time-domain effective-one-body model \SEOBNRvFIVEasym \citep{Estelles:2025zah}, asymmetries are incorporated through direct recalibration of the coprecessing waveform to precessing NR simulations.
Surrogate models such as \NRSurprec \citep{Varma:2019csw} include asymmetries inherently, as they are constructed directly from NR simulations of precessing systems that include all relevant physics.

In this work, we employ the quasicircular precessing BBH model in the frequency domain \IMRPhenomXPNR \citep{Hamilton:2025xru}, as implemented in LALSuite \citep{lalsuite}.
\IMRPhenomXPNR is based on \IMRPhenomXPHM \citep{Pratten:2020fqn,Garcia-Quiros:2020qpx,Pratten:2020ceb} and includes the $(\ell,|m|) = \{(2,2),(2,1),(3,3),(3,2),(4,4)\}$ multipoles in the co-precessing frame.
Precession dynamics during inspiral are described using the SpinTaylorT4 evolution \citep{Colleoni:2024knd}, while the merger-ringdown dynamics use a phenomenological description calibrated to NR simulations \citep{Hamilton:2021pkf,Thompson:2023ase}.
For the coprecessing waveform, the baseline aligned-spin model \IMRPhenomXHM \citep{Garcia-Quiros:2020qpx} is modified to incorporate calibration to precessing NR simulations in the late inspiral, merger and ringdown \citep{Hamilton:2021pkf}, and the model for asymmetries \citep{Ghosh:2023mhc}, both in the dominant multipoles.

\IMRPhenomXPNR is formulated in the frequency domain, whereas here we computed recoil using time-domain waveforms.
We therefore obtain time-domain signals via inverse Fourier Transform, applying a symmetric Tukey window with $\alpha=0.01$ prior to transformation.
Frequency-domain waveforms are generated with sufficiently fine frequency resolution to avoid spurious artifacts in the time domain.

\subsection{Recoil computation} \label{ssec:2.2}

The coalescence of a BBH system generically results in the emission of GWs carrying net linear momentum.
Conservation of momentum therefore implies that the merger remnant recoils relative to the center-of-mass frame \citep{Bekenstein:1973zz}, acquiring a velocity that can reach several thousand kilometers per second in extreme configurations \citep{Bruegmann:2007bri,Campanelli:2007cga,Lousto:2011kp,Lousto:2019lyf}.
The dominant contribution to the recoil arises during the late inspiral and merger phases of the signal, with larger recoils typically produced in systems with precessing spins.

A crucial requirement for reliable kick estimates is the inclusion of waveform asymmetries \citep[e.g.][]{Borchers:2024tdi}.
The equatorial symmetry of aligned-spin binaries enforces strong cancellations in the momentum flux, limiting recoil velocities to the orbital plane, while binaries that exhibit precession and broken equatorial symmetry, may have substantial out-of-plane kicks \citep{Borchers:2024tdi}.
Consequently, waveform models that do not incorporate equatorial asymmetries cannot reliably predict recoil velocities.

Early kick estimates were obtained from numerical relativity simulations \citep{Baker:2006vn}, and numerous fitting formulas \citep{Gonzalez:2006md,Healy:2018swt} and surrogate final state models \citep{Varma:2018aht,Varma:2019csw,Islam:2023mob,Islam:2025drw} have since been developed to estimate recoil velocities without full waveform generation.
The recent inclusion of waveform asymmetries in semi-analytic models has enabled the possibility of computing the recoil velocity analytically from the waveform \citep{Held:1980gc} using different waveform models.

Here we computed recoil velocities from the GW strain generated with \IMRPhenomXPNR, following the analytical expression for the radiated linear momentum $\vec{P}=(P_x,P_y,P_z)$ in e.g. \citet{Ruiz:2007yx},

\begin{subequations} \label{eq:kick_expression} \begin{align}
    P_\perp \equiv P_x + i P_y &= -\dfrac{1}{16\pi}\int_{-\infty}^\infty I_\perp(t)\  \mathrm{d}t,
    \\
    P_z &= -\dfrac{1}{16\pi}\int_{-\infty}^\infty I_z(t)\ \mathrm{d}t,
\end{align} \end{subequations}
where
\begin{subequations} \label{eq:kick_expression_2} \begin{align}
    I_\perp &= \sum_{\ell,m} \dot{h}_{\ell,m} \left(  a_{\ell,m}\dot{h}^\ast_{\ell,m+1} + b_{\ell,-m}\dot{h}^\ast_{\ell-1,m+1} - b_{\ell+1,m+1}\dot{h}^\ast_{\ell+1,m+1} \right),
    \\
    I_z &= \sum_{\ell,m} \dot{h}_{\ell,m} \left( c_{\ell,m}\dot{h}^\ast_{\ell,m} + d_{\ell,m}\dot{h}^\ast_{\ell-1,m} + d_{\ell+1,m}\dot{h}^\ast_{\ell+1,m} \right),
\end{align} \end{subequations}
with coefficients
\begin{subequations} \label{eq:kick_expression_coefficients} \begin{align}
    a_{\ell,m} &\equiv \dfrac{2\sqrt{(\ell-m)(\ell+m+1)}}{\ell(\ell+1)}, \\
    b_{\ell,m} &\equiv \dfrac{1}{\ell} \sqrt{\dfrac{(\ell-2)(\ell+2)(\ell+m)(\ell+m-1)}{(2\ell-1)(2\ell+1)}}, \\
    c_{\ell,m} &\equiv \dfrac{2m}{\ell(\ell+1)}, \\
    d_{\ell,m} &\equiv \dfrac{1}{\ell} \sqrt{\dfrac{(\ell-2)(\ell+2)(\ell-m)(\ell+m)}{(2\ell-1)(2\ell+1)}}
.\end{align} \end{subequations}

The dominant contribution to these integrals arises from the final few cycles prior to merger. Consequently, a shorter waveform starting at an initial frequency corresponding to the inspiral provides a sufficiently accurate approximation.
The total recoil velocity is then obtained by dividing the radiated momentum $\vec{P}$ by the remnant mass $m_f$, which can be obtained from the emitted GW energy via the waveform modes using an analogous expression \citep[see][for details]{Ruiz:2007yx}.
For the calculations presented here, we use the implementation provided in the \textsc{scri} package \citep{scri}.

Out-of-plane recoil components depend strongly on the azimuthal spin angles at merger, where waveform models exhibit systematic differences,
although the maximum and minimum kick values for a given configuration with varying in-plane spin angle are broadly consistent among waveform models \citep{Mielke:2024kya}.
As a result, kick estimates based on a single maximum-likelihood waveform realization can be highly model-dependent since even for a fixed set of masses and spin magnitudes and tilts, varying the in-plane spin orientation can produce a broad range of kick magnitudes.
In addition, spin angles are typically poorly constrained by GW observations \citep{Biscoveanu:2021nvg}, and their recovery exhibits waveform systematics \citep{Varma:2021csh}.
Then, to obtain statistically meaningful recoil predictions, we computed the kick for each posterior sample obtained from Bayesian parameter estimation. By averaging over the full distribution of intrinsic parameters, the strong dependence on poorly constrained angles is effectively marginalized and the impact of waveform systematics is minimized.
The resulting kick distributions provide a robust statistical characterization of the recoil imparted to each merger remnant.

\subsection{Parameter estimation}

Posterior samples used in this study were obtained following the inference methodology adopted for the \GWTCFOUR analysis presented in \citet{Xu:2025ajj}, using the \IMRPhenomXPNR waveform model.
In that work, GW signals are analyzed within a Bayesian framework, assuming stationary Gaussian noise and modeling the detector strain as the sum of noise and a coherent signal across the detector network.
Parameter inference is performed using the \dynesty nested-sampling package \citep{2020MNRAS.493.3132S}, as implemented in the \bilby inference framework \citep{Ashton:2018jfp,Romero-Shaw:2020owr}, with likelihood evaluations performed in the frequency domain.
The detector strain data, configuration files, power spectral densities, and calibration envelopes are taken directly from the \GWTCFOUR public release \citep{GWTC4release}
ensuring consistency with the LVK catalog analysis.

While all waveform models aim to describe the same underlying physical signal, differences in their physical assumptions and calibration can lead to systematic differences in the inferred posterior distributions for intrinsic parameters such as masses and spins.
Some parameters, such as the chirp mass and spin combinations such as the effective spin or the effective precessing spin, are more robustly constrained across waveform models, whereas quantities that depend on higher-order modes or precessional dynamics ---most notably the individual spin components, spin orientations, and derived parameters--- are subject to larger statistical and systematic uncertainties and should be interpreted with appropriate caution.

\section{BBH Environments} \label{sec:3}

Binary black hole mergers can originate in qualitatively different astrophysical environments, each leaving characteristic imprints on the distributions of intrinsic parameters.
In this section, we summarize the population models adopted in this work to represent the main formation channels: in dense stellar systems and isolated binary evolution in the field.
These synthetic catalogs provide the theoretical reference distributions against which we compare the intrinsic parameters inferred from GW observations in the following sections.

\subsection{Binaries in globular clusters} \label{ssec:3.1}

Dense stellar environments such as GCs provide a natural setting for the formation of BBHs.
Owing to their high stellar densities, black holes can efficiently form binaries through multi-body interactions, including three-body encounters.
Once formed, these binaries may undergo repeated dynamical interactions that harden the system and drive it toward merger.
These environments also allow for hierarchical mergers, in which the remnant of a previous coalescence is retained within the cluster and subsequently forms a new binary with another compact object \citep[see][and references therein for details]{2023arXiv231112118A}.

Metallicity is quantified through the relative abundance of iron to hydrogen, where the solar value is set at $Z_\odot=2\cdot10^{-2}$, as is the case in the population catalogs used here.
Cluster metallicity strongly influences the properties of BBHs that form within them. At low metallicity, stellar winds are weaker, producing more massive BHs \citep{Vink:2001cg,2010ApJ...714.1217B,Spera:2017fyx}.
In addition, metallicity correlates with formation epoch:
because heavy elements are synthesized through stellar evolution and supernova explosions, the early Universe was comparatively metal-poor.
As a result, low-metallicity GCs are generally associated with early formation times, whereas higher-metallicity clusters tend to form at later cosmic epochs.
Following the cluster formation history discussed by \citet{2019MNRAS.482.4528E}, we associate clusters with metallicities $Z/Z_\odot=10^{-2}$ and $10^{-1}$ with formation times of approximately $13$ and $12 \Gyr$ ago respectively, and near-solar metallicity clusters with formation times of roughly $2-5 \Gyr$ ago.

To model dense stellar environments, we use the CMC GC catalog \citep{Kremer:2019iul,2022ApJS..258...22R}, which includes 148 independent cluster simulations, generated by varying initial cluster parameters (total number of particles $N$, initial cluster virial radius $r_v$, metallicity $Z$ and galactocentric distance $R_{gc}$).
The simulations adopt the initial mass function of \cite{Kroupa:2000iv} with masses in the range of $0.08-150 \Msun$ and assume an initial stellar binary fraction of $f_b=5\%$.
The catalog uses the COSMIC code \citep{Breivik:2019lmt}, with SSE/BSE evolution codes \citep{Hurley:2000pk,Hurley:2002rf}, the rapid SN model for remnant treatment from \citet{2012ApJ...749...91F} and metallicity dependent winds from \citet{Vink:2001cg} \citep[see Sect. 2.1. in][for more details]{Kremer:2019iul}. 
For many decades, GCs have been regarded as large gravitationally bound collections of stars characterized by a single formation epoch and a nearly uniform chemical composition. 
However, recent observations have established that GCs contain multiple stellar populations with significant variations in light element abundances \citep{2019A&ARv..27....8G,2018ARA&A..56...83B,2022Univ....8..359M}.
In the simulations used to construct the CMC catalog, all stars are however assumed to form at a fixed time with a single metallicity, thereby bypassing the complex early phases of cluster and stellar formation associated with molecular cloud collapse.

The CMC catalog provides intrinsic properties of compact-object binaries formed within the simulated clusters, including estimates of the cluster escape velocity.
Ongoing dynamical interactions can impart velocities to binaries that exceed the local escape speed, leading to their ejection from the cluster prior to merger.
Typical escape velocities for GCs are of order tens of $\kms$, with values depending on the cluster mass, concentration, and evolutionary state.
In this work, we use the escape velocities predicted by the CMC catalog to check and validate observational estimates, and to assess the retention probability of merger remnants in dense environments.
This allows us to connect recoil velocity distributions derived from GW observations to the likelihood of hierarchical mergers occurring within GCs.

\subsection{Field binaries} \label{ssec:3.2}

Binary black holes may also form and evolve in isolation, with no dynamical interactions, and the binary evolution is driven primarily by stellar evolution processes.
This population, commonly referred to as \emph{field binaries}, originates from isolated stellar binaries \citep[see][and references therein]{Mandel:2021smh}.
If the binaries stay bound throughout their evolution, they eventually form compact-object binaries following successive core-collapse events.
In contrast to dense stellar environments, field binaries are not expected to undergo dynamical exchanges of hierarchical mergers, and their properties reflect only the cumulative effects of mass transfer, common-envelope evolution, and supernova explosions.

Here we model the field population using the cosmological binary mergers catalog of \citet{Olejak:2022zee}, generated with the StarTrack population synthesis code \citep{2002ApJ...572..407B,2008ApJS..174..223B,2021A&A...651A.100O}.
These simulations adopt two different prescriptions for the pair-instability supernovae (PSN): (i) a strong PSN which limits the BH masses to $\sim 45 \Msun$ as adopted in \cite{2016A&A...594A..97B}, and (ii) a revised prescription from \cite{2020ApJ...905L..15B} in which stars experience disruption in PSN if their final He core mass lies in $M_\mathrm{He} \in [90, 175] \Msun$. This revised model does not include any mass loss in pulsation PSN and allows formation of BHs with masses up to $90 \Msun$.
In addition to this, binary evolution is modeled using two different prescriptions for stability of Roche Lobe Overflow (RLOF):
(i) the standard common-envelope (CE) phase development criteria \citep[see][for details]{2008ApJS..174..223B}, and (ii) a revised mass transfer stability criteria \citep{2017MNRAS.465.2092P,2021A&A...651A.100O}.
These simulations use the remnant mass prescriptions (NSs and BHs) given by \cite{2022ApJ...931...94F}, using a convective-engine model driving core-collapse supernovae.
For our analysis we use models with the revised PSN prescription, revised mass-transfer stability criteria, and three values for the convection mixing parameter $f_\mathrm{mix}=0.5,1.0,4.0$. This parameter is inversely proportional to convection growth time, and $f_\mathrm{mix}=4.0$ corresponds to a rapid growth of the convection in $\lesssim10\ms$ which develops into an explosion in the first $\sim 100\ms$.
In Appendix~\ref{app:A} we repeat the analysis using the revised PSN prescription and the same three $f_\mathrm{mix}$ values, but adopting the standard CE prescription.

BHs formed from stellar collapse while they are part of field binaries are expected to have similar properties to first-generation cluster BHs,
but their subsequent evolution differs substantially.
BBH formation in the field proceeds through the evolution of primordial stellar binaries, and BH spins may be influenced by mass accretion during stable mass transfer or common-envelope phases, rather than by dynamical capture or repeated mergers.
As a result, spin magnitudes in field binaries are moderate \citep[see e.g.][]{Xu:2025sqj} and spin orientations are expected to be more closely aligned with the orbital angular momentum, and large spin misalignments are less common than in \revised{dense-cluster formed} systems.
This does not negate the existence of primordial stellar binaries in clusters (CMC catalog has $f_b=5\%$), with quick evolution that mitigates the potential impact of cluster dynamics.

While quantitative predictions vary across models, field binaries are generally expected to dominate the BBH merger rate in the local Universe, with coalescence rates typically exceeding those of globular clusters by one to two orders of magnitude \citep{Mandel:2021smh}.
This makes the field population an essential baseline when assessing the contribution of dense stellar environments.

\subsection{Other gravitationally bound environments}

Nuclear star clusters (NSCs) represent another class of dense stellar environments capable of forming BBHs.
Compared to globular clusters, NSCs are more massive and can have escape velocities of several hundred $\kms$ \citep{Gerosa:2020bjb,2019MNRAS.486.5008A,2020MNRAS.498.4591F}.
As a result, merger remnants are more likely to be retained than in GCs, increasing the likelihood of hierarchical mergers and the formation of very massive BHs.

\revised{Although our population-classification framework is based on globular-cluster simulations, in Sect.~\ref{sec:5} we consider NSCs in the context of remnant retention because their larger escape velocities make them more natural environments for hierarchical growth.
We stress, however, that NSCs are not simply scaled-up globular clusters: their formation histories, metallicity distributions, stellar populations, and dynamical evolution can differ substantially.
The GC catalogs adopted here should therefore not be interpreted as dedicated NSC population models.}

Beyond star clusters, merger remnants may also remain bound within their host galaxies.
Typical galactic escape velocities exceed those of globular clusters \citep{Merritt:2004xa} and highly depend on distance to the galactic centre.
Evaluating recoil velocities relative to galactic escape speeds allows us to also distinguish between retention within galaxies and complete ejection into intergalactic space.

\section{Environment of detected BBHs} \label{sec:4}

In this section we assess whether individual GW events are more consistent with formation in dense stellar environments or in the field.
We first constructed detectable subsets of the population catalogs, then defined the parameter spaces used for comparison, and finally quantified the preference of each event through Bayes factors.

\subsection{Preparing the catalogs} \label{ssec:4.1}

To assess the likely formation environments of BBH events,
we compared their inferred intrinsic parameters with synthetic populations of BBHs formed in dense stellar environments \citep{Kremer:2019iul,2022ApJS..258...22R} and in the field \citep{Olejak:2022zee}.
These catalogs provide complementary predictions for BBH populations formed in clusters and through isolated binary evolution in sparse environments.

A direct comparison between observed GW events and population catalogs must account for detectability of binaries, which is strongly linked to detector sensitivity.
The \GWTCFOUR population analysis by the LVK Collaboration \citep{LIGOScientific:2025pvj} finds that $99\%$ of detectable BBHs lie at redshifts $z\leq 1.5^{+0.2}_{-0.2}$.
We therefore restricted both catalogs to mergers occurring at $z<1.5$, ensuring consistency with the observed sample.
This cut is a simple way of accounting for the local portion of the Universe that is LVK-observable, \revised{although it could still include population regions that are hardly detectable by current detectors. This cut is only an approximate proxy for detectability, and could be refined using more sophisticated detectability-weighted methods as suggested in \citet{Mould:2023ift}, accounting for intrinsic parameters, orientation, and detector sensitivity. Consequently, the resulting populations should be interpreted as conditional on this treatment.}
\revised{To assess the robustness of our conclusions against selection effects, we considered a different catalog cut, based on the SNR of the signals emitted by the binaries, estimated using the \gwfish package \citep{Dupletsa:2022scg} for signals computed with \IMRPhenomXPNR in a 3-detector (LHV) network, randomly oriented and located in the sky. Then, we compared the distributions of the 6 populations on the 3 parameter spaces cuts at different SNRs to their corresponding versions pruned at $z<1.5$. We validate that the JS divergences of the redshift-cut distributions are of at $\lesssim0.05$ when considering a cut at $\mathrm{SNR}>3$.}

This limitation is particularly relevant for dense stellar environments, whose metallicities trace cosmic time,
as clusters formed at early epochs may host BBHs that merge at high redshift, and are therefore unobservable by the current detector network.
Such low-metallicity clusters are therefore more depleted of massive BHs than younger clusters.
Ignoring this effect can bias comparisons between observed events and population catalogs.

\begin{figure}
    \centering
    \includegraphics[width=\linewidth]{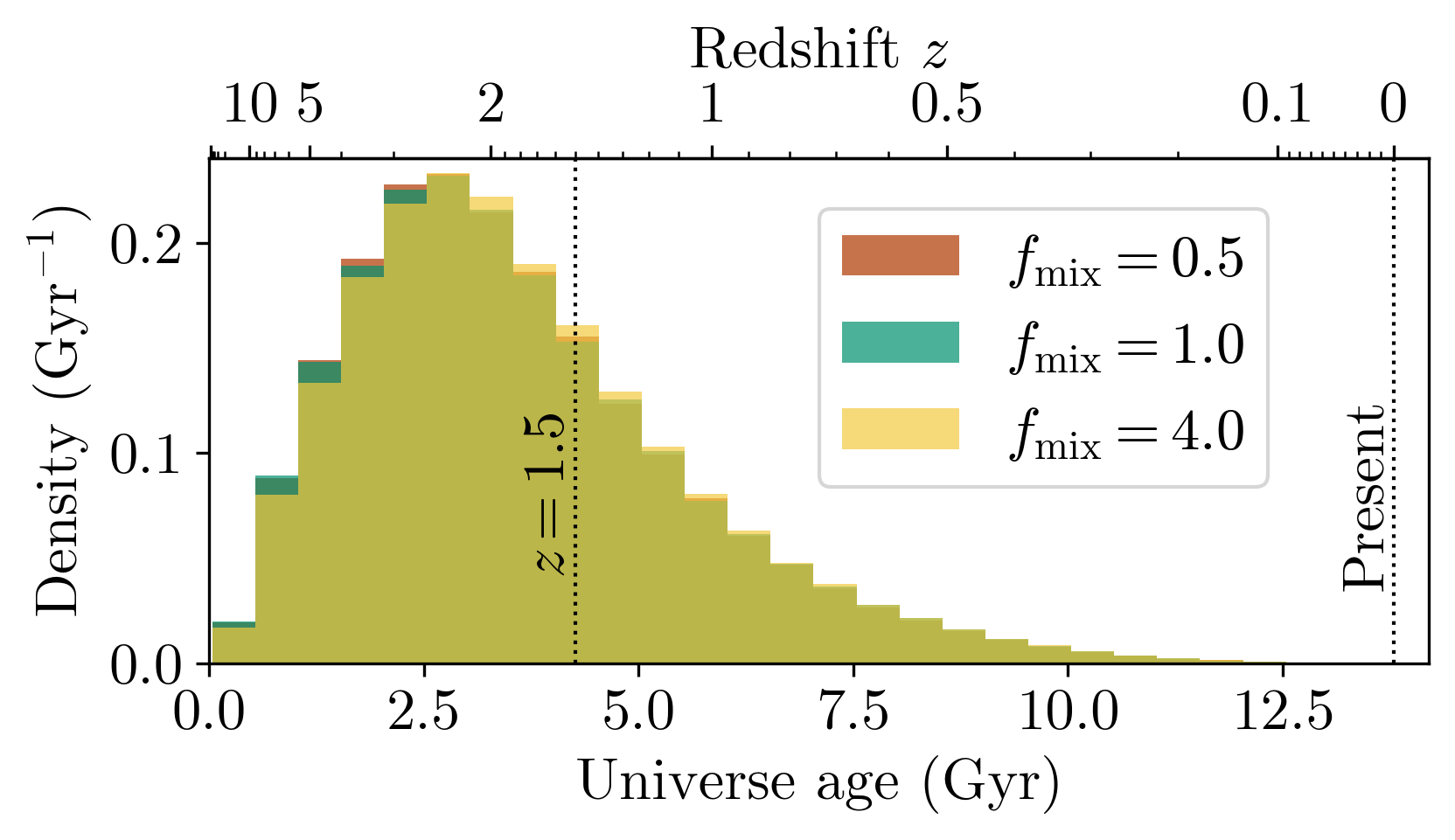}
    \caption{Merger time distribution of field binaries for populations of field binaries with different convection mixing parameters $f_\mathrm{mix}$. In this work, we only consider binaries at the RHS of the $z=1.5$ line.}
    \label{fig:1}
\end{figure}

For the field catalog \citep{Olejak:2022zee}, the redshift at merger is provided for each simulated system, so we simply retained binaries merging within $z<1.5$.
Fig.~\ref{fig:1} shows the merger time distributions for field populations with different $f_\mathrm{mix}$, together with the $z=1.5$ threshold. Around $30\%$ of the $10^6$ mergers in each catalog fall within the observable window.

For the GC catalog, the procedure is more involved.
The CMC simulations provide the delay time between star formation and binary merger, as well as whether the merger occurs inside or outside the host cluster. To convert merger delay times into merger redshifts, we used the formation time for the host cluster stated in Sect.~\ref{ssec:3.1}, from \citet{2019MNRAS.482.4528E}.
Using these formation times, we identify BBHs whose merger times correspond to $z<1.5$.

An additional complication arises because not all BBHs formed in GCs merge within the cluster potential. Dynamical encounters can eject binaries prior to merger, leading them to coalesce in the field.
Since our goal is to assess remnant retention in dense environments, we retained only binaries that merge inside their host clusters and satisfy the redshift cut.

\begin{figure}
    \centering
    \begin{subfigure}{\columnwidth}
        \centering
        \includegraphics[width=\linewidth]{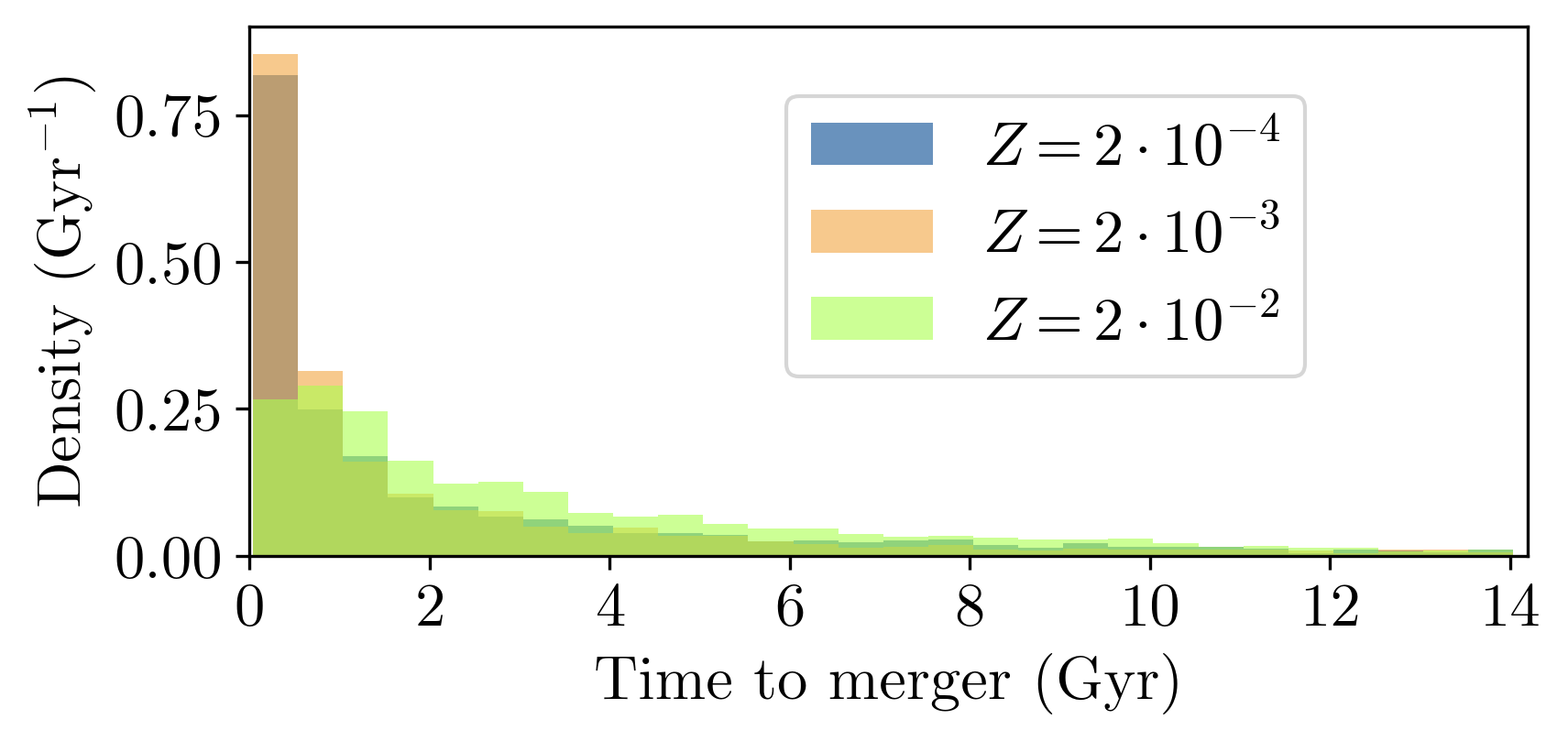}
        \caption{Distribution of merger delay time, or time from cluster formation to merger, for BBHs in GCs of different metallicities. BHs in lower-metallicity clusters tend to be more massive and therefore merge earlier.}
        \label{fig:2a}
    \end{subfigure}
    \begin{subfigure}{\columnwidth}
        \centering
        \includegraphics[width=\linewidth]{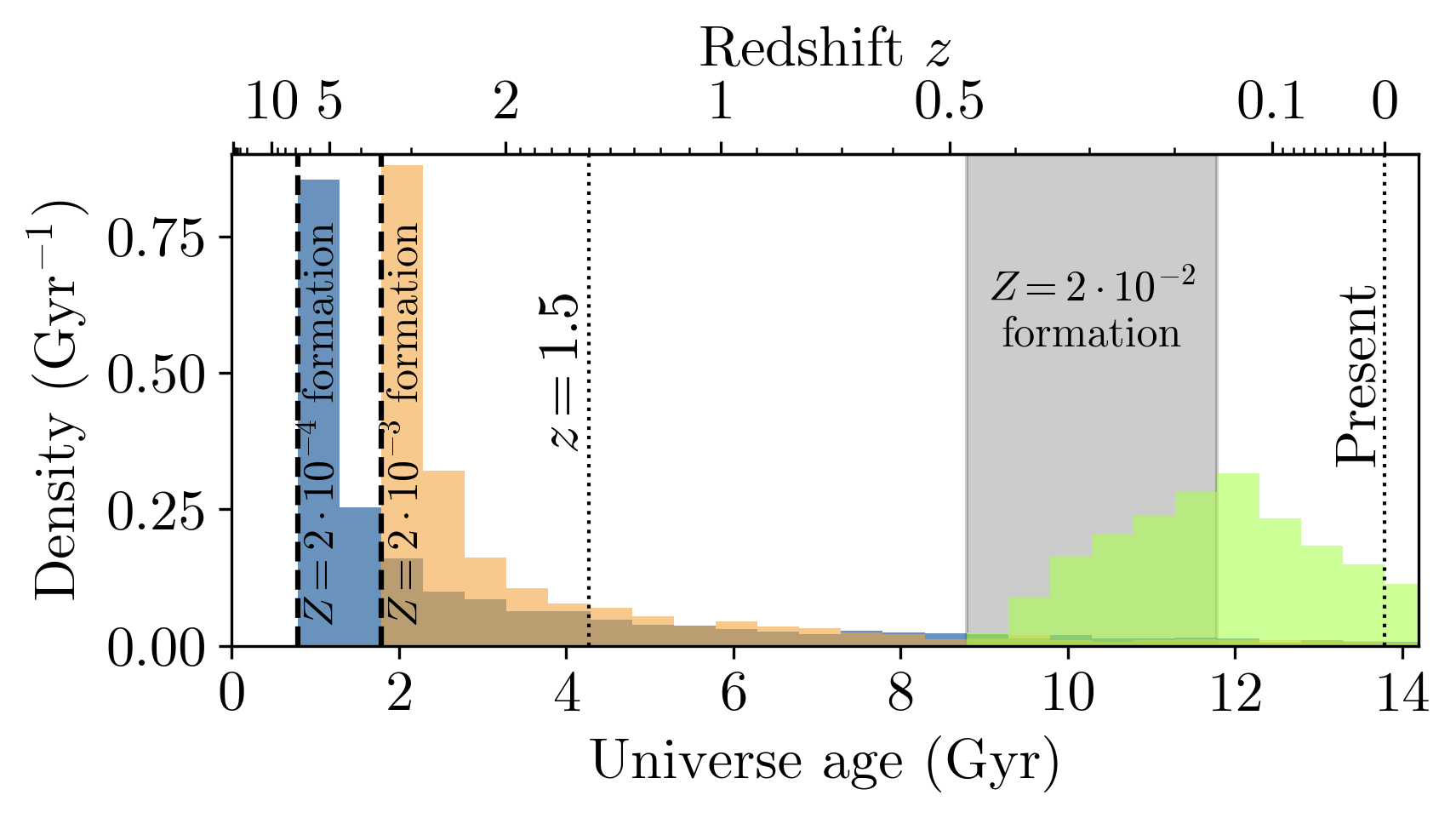}
        \caption{Merger time distribution for BBHs in GCs of different metallicities. The formation time has been taken into account and therefore the top panel distributions are shifted.}
        \label{fig:2b}
    \end{subfigure}
    \caption{Merger time distribution for GC catalogs. Only the binaries at the RHS of the $z<1.5$ line pass the cut. These figures contain all binaries in the catalogs, not only those merging within their host clusters.}
    \label{fig:2}
\end{figure}

This selection produces a detectable subset of the full CMC catalog.
In general, low-metallicity clusters form more massive BBHs, which merge more rapidly, thus many BBHs merge at high redshift and are excluded by our selection.
Figure~\ref{fig:2} shows the distribution of merger delay times for clusters of different metallicities.
The top panel displays the delay time between cluster formation and merger. When cluster formation epochs are incorporated, the distributions shift in cosmic time (bottom panel), allowing identification of mergers with the detectable window of $z<1.5$.

\begin{figure}
    \includegraphics[width=\columnwidth]{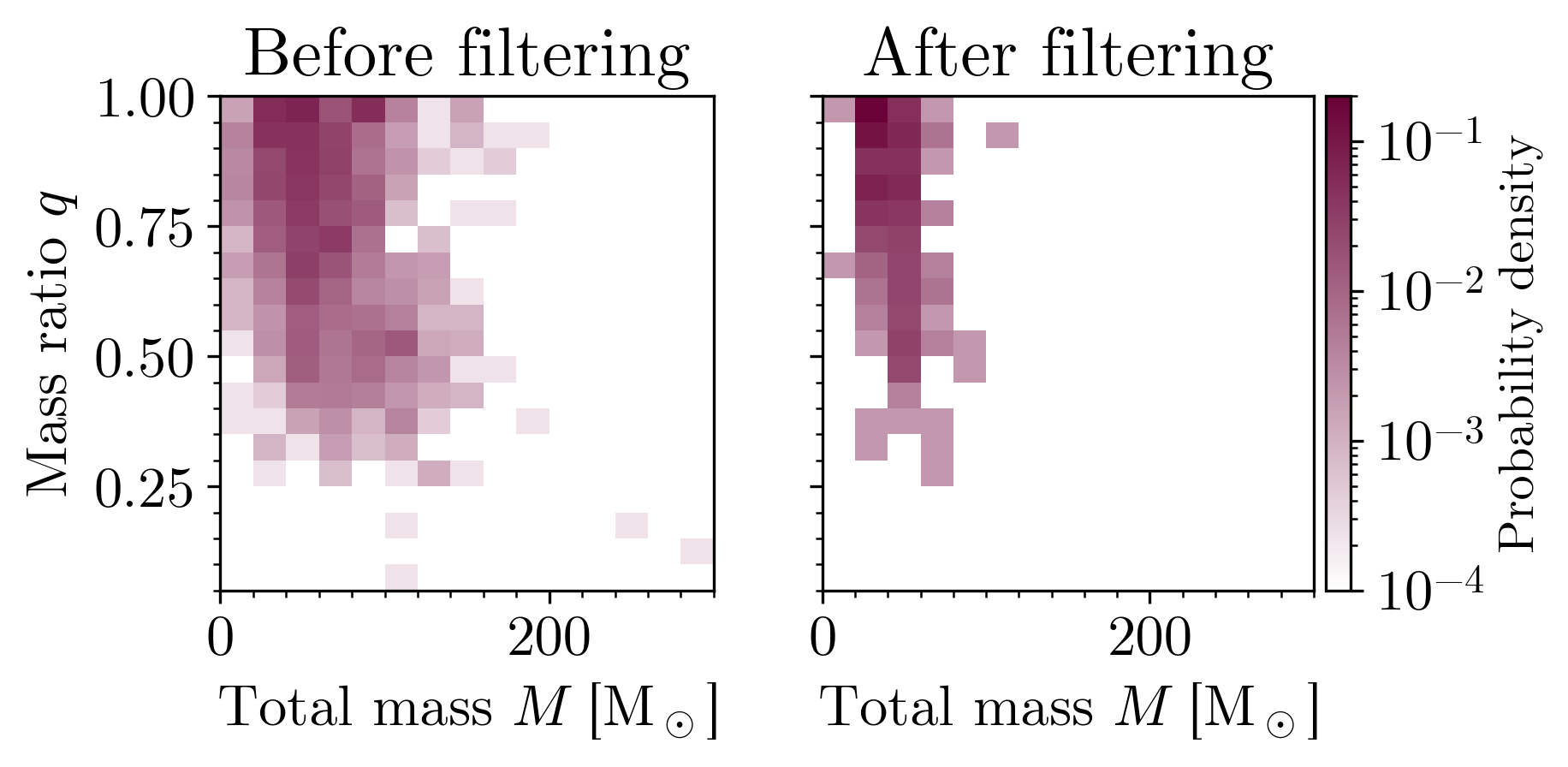}
    \caption{Mass distribution of binaries in the $Z=2\cdot10^{-4}$ clusters. The left panel includes the 4333 binaries in the catalog with that metallicity value and the right panel includes only the 460 that are selected. The pruning due to imposing the $z\leq 1.5$ redshift cut is significant, specially at higher masses.}
    \label{fig:3}
\end{figure}

Figure~\ref{fig:3} illustrates how the BBH mass distributions for low-metallicity clusters changes after imposing the $z<1.5$ cut.
The removal of high-redshift mergers significantly reduces the high-mass tail of the distribution.
A small shift in the formation times does not affect the distributions deeply.
Table~\ref{tab:1} summarizes the number of first-generation and higher-generation mergers remaining after successive selections.
Table~\ref{tab:2} reports the characteristic escape velocities $\vesc$ of host clusters in the CMC catalog. These values, typically in the range $\sim 50-100\kms$, are consistent with observationally inferred values from electromagnetic observations used in the retention analysis of Sect.~\ref{ssec:5.2}.

\begin{table*}[t]
    \centering
    \begin{tabular}{cccc}
    \hline \hline
     & $Z=2\cdot10^{-4}$ & $Z=2\cdot10^{-3}$ & $Z=2\cdot10^{-2}$ \\ \hline
    All binaries in the catalog             & $4333$ ($3674+659$)   & $4206$ ($3580+626$)   & $4581$ ($3648+933$) \\
    Binaries that merge inside the cluster  & $2229$ ($1893+336$)   & $2042$ ($1722+320$)   & $2572$ ($2016+556$) \\
    Binaries that merge within $z<1.5$      & $460$ ($403+57$)  & $452$ ($403+49$)  & $1644$ ($1288+356$) \\ \hline
    \end{tabular}
    \caption{Binary populations in the CMC catalog, separated by metallicity. Number of 1st generation (1G) and higher-generation (HG) mergers are provided, as ``Total (1G+HG)''.}
    \label{tab:1}
\end{table*}

\begin{table}
    \centering
    \begin{tabular}{cccc}
    \hline \hline
    Generation & $Z=2\cdot10^{-4}$ & $Z=2\cdot10^{-3}$ & $Z=2\cdot10^{-2}$ \\ \hline
    Inside cluster & $55^{+63}_{-32}\kms$ & $51^{+59}_{-30}\kms$ & $57^{+43}_{-34}\kms$ \\
    Within $z<1.5$ & $38^{+26}_{-19}\kms$ & $36^{+15}_{-19}\kms$ & $67^{+36}_{-42}\kms$ \\ \hline
    \end{tabular}
    \caption{$\vesc$ median value and $90\%$ symmetric confidence interval for binaries in the CMC catalog, as provided by the catalog.}
    \label{tab:2}
\end{table}

This pruning procedure ensures that both the cluster and field catalogs used in this work represent BBH populations \revised{within the redshift range accessible to} the LVK detectors during \OFOURa,  \revised{making them comparable to the detected events in this respect}.

\subsection{Parameters}

To assess whether a given event shows more overlap with the adopted dense-cluster populations or in the field, we compare its intrinsic parameters inferred from GW PE with the distributions predicted by the population catalogs.
The choice of parameters to compare is crucial: they must be astrophysically informative, reasonably well constrained by GW observations, and available in the population synthesis catalogs.

The component masses $m_1$~and~$m_2$ are among the best-measured intrinsic parameters in GW observations and carry direct astrophysical information about the formation channel.
Although the total mass $M=m_1+m_2$ primarily acts as a scaling parameter in the GW signal it is typically well constrained when the merger contributes to high signal-to-noise ratio, it is highly relevant for population studies.
By contrast, the chirp mass $\mathcal{M}$, although typically very tightly constrained in PE specially for low-mass systems due to its leading order impact on the frequency during the inspiral, does not have a direct astrophysical interpretation outside of the context of GW analysis.
For this reason, we worked with the pair $(M,q)$.

Spin information can provide additional insight into formation channels.
Individual spin magnitudes $\chi_i$ are generally weakly constrained by PE, particularly for moderate signal-to-noise ratio events.
To mitigate this limitation, we considered instead the effective inspiral spin parameter \citep{Ajith:2009bn,Santamaria:2010yb},
\begin{equation}
    \chieff = \dfrac{m_1 \chionepar + m_2 \chitwopar}{M}
,\end{equation}
which is often better measured and encodes the spin components aligned with the orbital angular momentum.
This parameter therefore captures information from both spin magnitudes and orientations, which differ across formation channels.

We could also compute the effective precessing spin parameter \citep{Schmidt:2014iyl},
\begin{equation}
    \chip = \max{\left( \chioneperp,\ \dfrac{4q^2+3q}{4+3q} \, \chitwoperp\right)}
,\end{equation}
which captures the dominant in-plane spin effects.
However, $\chip$ is generally poorly constrained in GW posterior distributions, so we do not use it to infer population preferences.

The treatment of spins in the cluster population requires additional assumptions.
The CMC simulations provide individual spin magnitudes but do not track spin orientations. As a result, $\chieff$ cannot be computed directly without specifying angular distributions.
We assumed isotropic spin orientation in dense environments, consistent with expectations for dynamically assembled binaries.
Therefore, we randomly sample spin directions before computing $\chieff$.

The cluster spin-magnitude distribution in the catalogs is sharply structured:
first-generation BHs are typically non-spinning \citep{Fuller:2019sxi}, while higher-generation BHs have spin magnitudes accumulated around $\chi_i\simeq 0.69$, which coincides with the remnant spin magnitude of a non-spinning equal-mass system \citep{Barausse:2009uz}.
This behavior is astrophysically expected, since significant spin-up or spin-down in dense environments is not expected due to accretion, before the moment a compact object interacts in an N-body encounter or becomes gravitationally attached in a binary.
The spin magnitude of higher-generation mergers in the population synthesis catalogs coincides with the expected remnant spin magnitude of non-spinning equal-mass BHs \citep[see e.g.][]{Healy:2014yta}. However, the distribution for unequal-mass binaries is less sharp and peaks at around $0.75$. \citep{Lousto:2009ka,PhysRevD.82.129902}.

For each population realization (varying $Z$ in clusters and $f_\mathrm{mix}$ in the field) we constructed multidimensional probability density functions $\pi(\mathbf{x}|\mathrm{pop})$ using Kernel Density Estimation (KDE).
Because the cluster populations contain only a few hundred binaries after the selection described in Sect.~\ref{ssec:4.1}, constructing KDEs in the full three-dimensional parameter space $\mathbf{x}=(M,q,\chieff)$ could be statistically fragile, with a risk of overfitting or oversmoothing.
We therefore restricted our analysis to two-dimensional subspaces: $\mathbf{x} = (M,q)$, $\mathbf{x} = (M,\chieff)$ and $\mathbf{x} = (q,\chieff)$. This choice could be extended to include additional parameters, such as $\chip$, or replaced with alternative approaches, such as adaptive KDEs \citep{Sadiq:2025vly}, that better handle higher-dimensional spaces, or normalizing flows \citep[see][for recent uses]{Colloms:2025cjx,Scarpa:2026piy}.

The KDEs were constructed \revised{directly in the variables $M$, $q$, and $\chieff$, without logarithmic transformations.} 
\revised{Different kernel choices or variable transformations may quantitatively affect the resulting Bayes factors.}
We use Gaussian kernels with automatic bandwidth determination, as implemented in \texttt{scipy.stats.gaussian\_kde} \citep{scipy}, and were built from the simulated binaries that satisfy the selection criteria described in Sect.~\ref{ssec:4.1}.

\begin{figure}[b]
    \centering
    \begin{subfigure}{\columnwidth}
        \centering
        \includegraphics[width=0.9\linewidth]{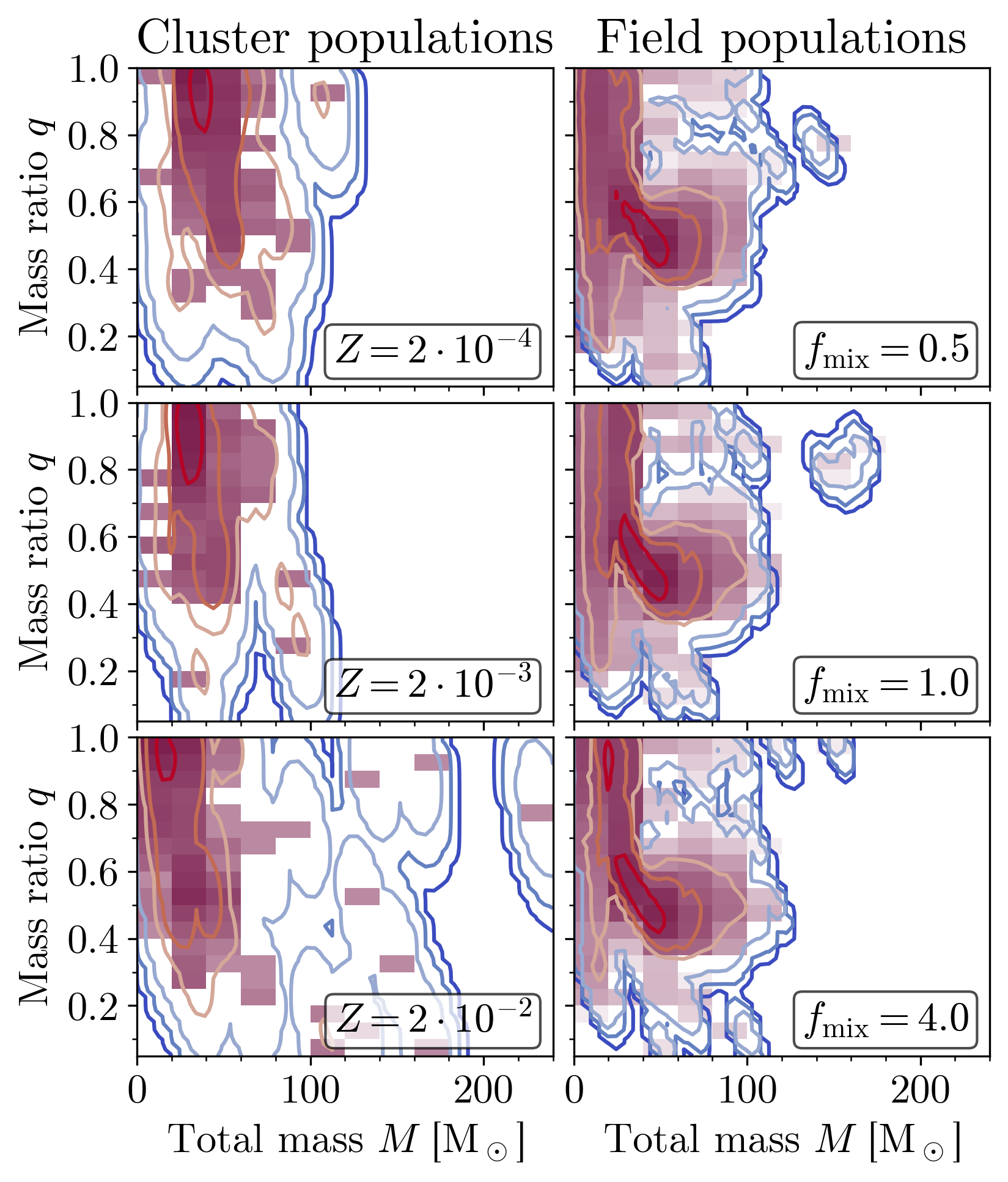}
        \caption{Population distribution for $\mathbf{x}=(M,q)$.}
        \label{subfig:4a}
    \end{subfigure}
\end{figure}

\begin{figure}
    \ContinuedFloat
    \begin{subfigure}{\columnwidth}
        \centering
        \includegraphics[width=0.9\linewidth]{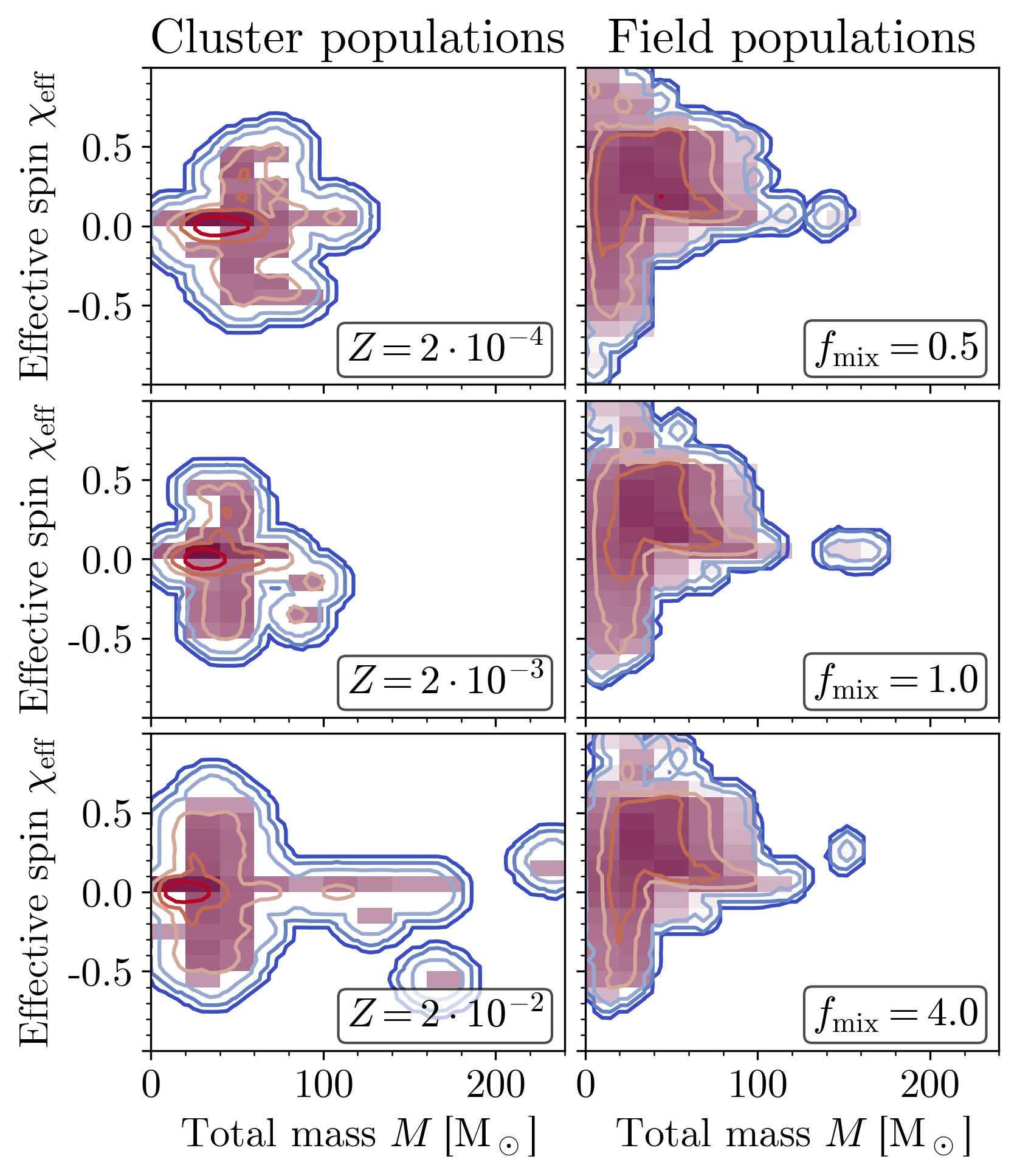}
        \caption{Population distribution for $\mathbf{x}=(M,\chieff)$.}
    \end{subfigure}
    \begin{subfigure}{\columnwidth}
        \centering
        \includegraphics[width=0.9\linewidth]{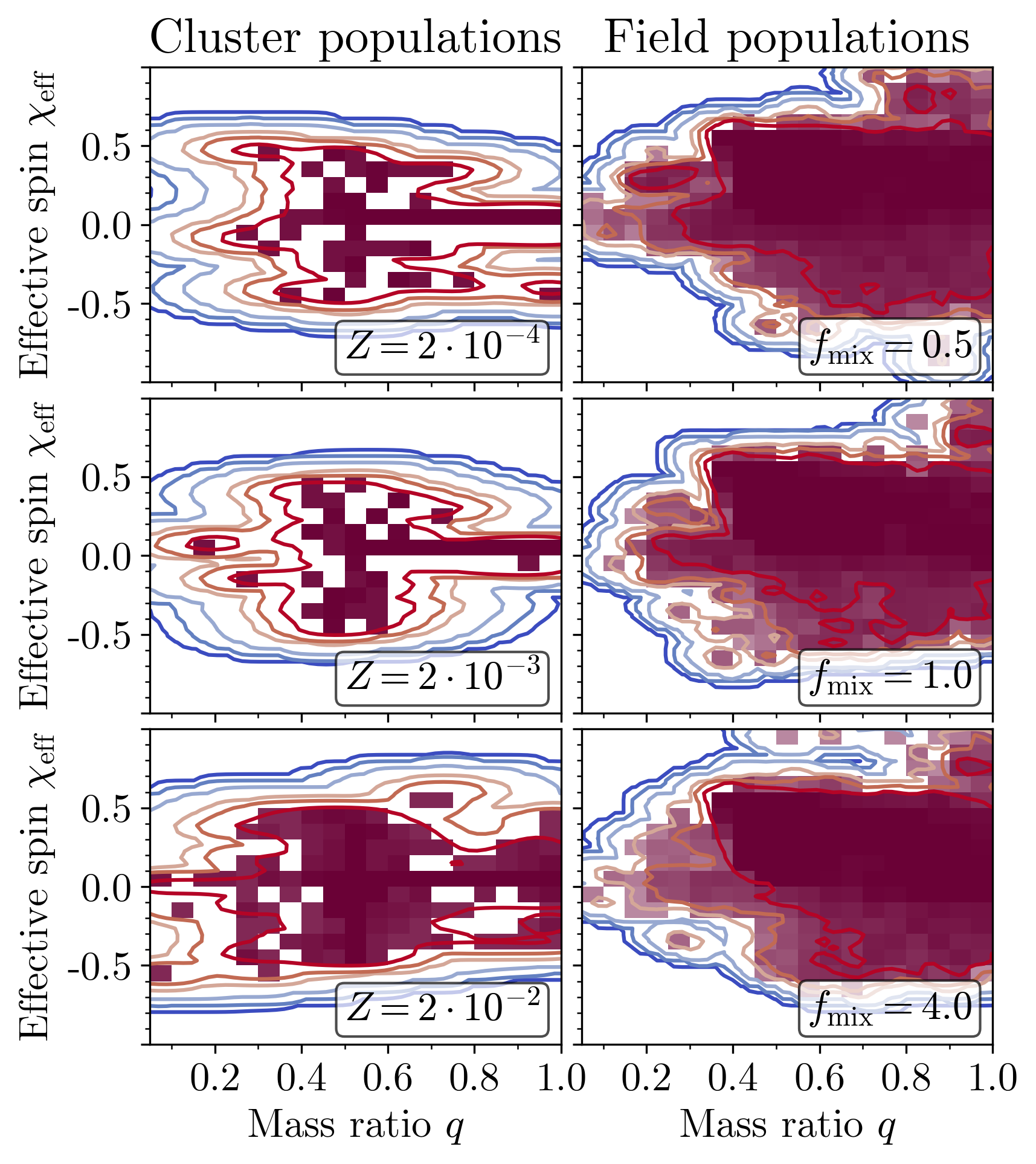}
        \caption{Population distribution for $\mathbf{x}=(q,\chieff)$.}
    \end{subfigure}
    \begin{subfigure}{\columnwidth}
        \centering
        \includegraphics[width=0.5\linewidth]{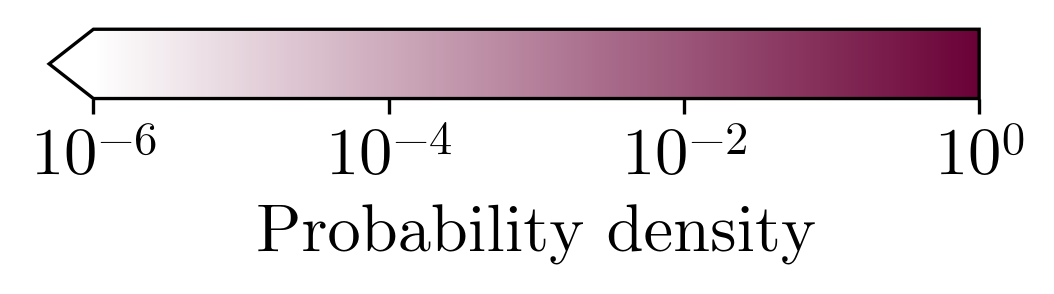}
        \includegraphics[width=0.45\linewidth]{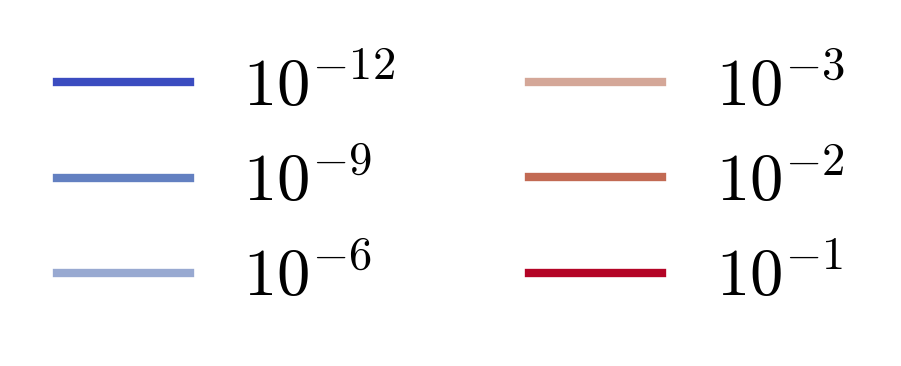}
        \caption{Colorbar and contour line legend for all subfigures.}
    \end{subfigure}
    \caption{Population distributions are shown, both as two-dimensional histograms and kernel density estimates (KDEs) overlaid as contour lines.
    In each subfigure corresponding a different parameter subspace, cluster binaries with varying metallicity are displayed in the left column, while field binaries with different $f_\mathrm{mix}$ are shown in the right column.
    The numbers labeling the contours correspond to the numerical value of the KDE.}
    \label{fig:4}
\end{figure}

Figure~\ref{fig:4} shows the resulting population distributions for cluster and field binaries in the considered two-dimensional parameter subspaces.
For each subfigure representing a subspace, the three cluster populations are shown on the left column and the three field populations in the right column, both as two-dimensional histograms and KDE contours.

Clear differences emerge between the formation channels.
In general, \revised{the adopted} cluster populations extend to higher total masses and tend to present more equal-mass systems, particularly at low metallicity, while field binaries occupy a narrower mass range.
As for $\chieff$, cluster populations reflect the assumed isotropic orientations and characteristic spin magnitudes of first- and higher-generation BHs.
Field binaries typically show a preference for $\chieff>0$, with broader ranges, specially at lower masses and near-equal mass ratios. 

These complementary differences in mass and spin distributions provide the basis for the statistical comparison performed in the following section, where we compute Bayes factors for individual GW events, quantifying their relative preference for a dense or sparse formation environment.

\subsection{Bayes factor of an event} \label{ssec:4.3}

After constructing population probability density functions for BBHs formed in dense stellar environments and in the field, we quantified how consistent individual events are with each formation channel.
Rather than identifying specific ``smoking-gun'' signatures such as extreme masses or spins, we adopted a population-based approach that compares each event against the full parameter distributions predicted by the catalogs.

\begin{figure*}
    \centering
    \includegraphics[width=1.9\columnwidth]{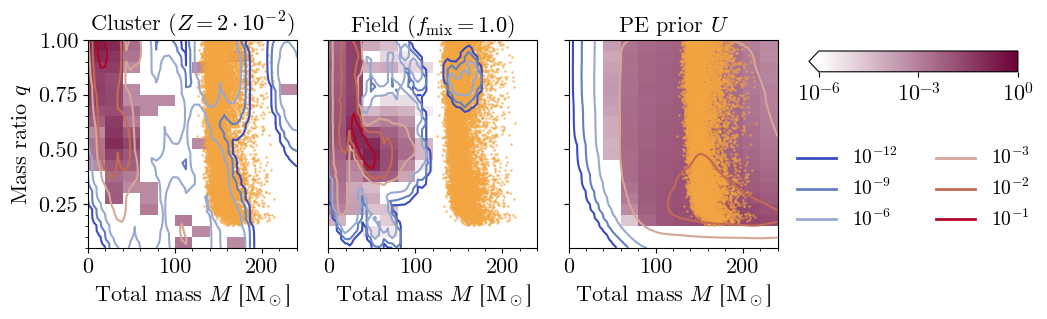}
    \caption{Population distribution for the subspace $\mathbf{x}=(M,q)$ and event \fullevent{GW231028}.
    Left and center panels show the population of cluster (with $Z=2\cdot10^{-2}$) and field (with $f_\mathrm{mix}=1.0$) binaries.
    Right panel shows the PE prior distribution for that event.
    In all panels, the PE posterior samples are shown.
    $\log_{10}w_\mathrm{pop}=-1.78$ ($-3.80$) for this cluster (field) population.}
    \label{fig:5}
\end{figure*}

Let \emph{pop} be a population model characterized by a probability density function $\pi(\mathbf{x}\,|\,\mathrm{pop})$, then the evidence for an event data $\mathbf{d}$ under such population prior is
\begin{equation} \label{eq:popevidence}
    \mathcal{Z}(\mathbf{d}\,|\,\mathrm{pop}) = \int \mathcal{L}(\mathbf{d}\,|\,\mathbf{x})\ \pi(\mathbf{x}\,|\,\mathrm{pop})\ \mathrm{d}\mathbf{x}
,\end{equation}
where $\pi(\mathbf{x}\,|\,\mathrm{pop})$ is the parameter distribution of the model and $\mathcal{L}(\mathbf{d}\,|\,\mathbf{x})$ is the likelihood of the data.
The posterior distributions are computed using a prior $U$ set in the analysis by the LVK Collaboration. The KDE for the prior $\pi(\mathbf{x}|U)$ is computed in the same way as for the astrophysical populations.

Using Bayes theorem for posterior samples
\begin{equation}
    p(\mathbf{x}\,|\,\mathbf{d},U) = \dfrac{ \mathcal{L}(\mathbf{d}\,|\,\mathbf{x})\ \pi(\mathbf{x}\,|\,U) }{\mathcal{Z}(\mathbf{d}\,|\,U)}
,\end{equation}
Equation~\eqref{eq:popevidence} can be rewritten as
\begin{equation}
    \mathcal{Z}(\mathbf{d}\,|\,\mathrm{pop}) = \mathcal{Z}(\mathbf{d}\,|\,U)\int \dfrac{\pi(\mathbf{x}\,|\,\mathrm{pop})}{\pi(\mathbf{x}\,|\,U)}\ p(\mathbf{x}\,|\,\mathbf{d},U)\ \mathrm{d}\mathbf{x},
\end{equation}
where the posterior samples have been reweighted.
As pointed out by e.g. \citet{Mould:2023ift}, not reweighting the posterior samples would introduce a bias in the computation.
Since we were only interested in relative population weights, we can factor out $\mathcal{Z}(\mathbf{d}\,|\,U)$. In practice, we use the quantity
\begin{equation}
    w_\mathrm{pop} = \dfrac{\mathcal{Z}(\mathbf{d}\,|\,\mathrm{pop})}{\mathcal{Z}(\mathbf{d}\,|\,U)} \cong \dfrac{1}{N_\mathrm{samp}} \sum_{i=1}^{N_\mathrm{samp}} \dfrac{\pi(\mathbf{x_i}\,|\,\mathrm{pop})}{\pi(\mathbf{x_i}\,|\,U)}
,\end{equation}
where $\{\mathbf{x_i}\}_{i=1}^{N_\mathrm{samp}}$ are the posterior PE samples and we have substituted the integral for a summation over these samples.

Figure~\ref{fig:5} illustrates this procedure for the event \fullevent{GW231028}, for $\mathbf{x}=(M,q)$.
The PE posterior samples $\{\mathbf{x_i}\}_{i=1}^{N_\mathrm{samp}}$ are overlaid on the representative cluster and field population distributions, and the corresponding values of $w_\mathrm{pop}$ are indicated.
In this example, the overlap with this particular cluster population is larger than with the chosen field population.
However, this alone is not definitive of preference for any origin, as we need a procedure that accounts for all possible realizations of cluster and field populations to draw conclusions.

To compare the cluster and field populations directly, we define the Bayes factor
\begin{equation} \label{eq:odds_ratio}
    \bayes(\mathrm{event}) = \dfrac{w_\mathrm{cluster}(\mathbf{d})}{w_\mathrm{field}(\mathbf{d})}
\,.\end{equation}

For dense stellar environments, multiple cluster populations are available, with varying metallicity $Z$.
Similarly, the field catalog contains several realizations corresponding to different assumptions about binary evolution, characterized by different values of $f_\mathrm{mix}$.
We need to consider Bayes factors for all combinations of cluster and field realizations.

In this study, we have not committed to a specific realization of either formation channel.
Instead, we remained agnostic about the underlying assumptions entering the population models and required consistency across all available realizations.
Accordingly, for a cluster origin to be favored, the event must be preferred for at least one cluster population at a given metallicity when compared to all field populations.
This criterion ensures that any preference for cluster formation is not driven by a particular choice of parameters, but reflects a robust distinction between the two formation channels.

\revised{The quantity $\mathcal{B}_\mathrm{c/f}$ measures the relative consistency of an event with the adopted cluster and field population models, but does not by itself represent an astrophysical posterior probability for either formation channel.
Interpreting these Bayes factors as posterior odds requires additional assumptions about the relative merger rates of the two channels,
\begin{equation}
    \mathcal{O}_\mathrm{c/f} = \bayes\ \dfrac{R_\mathrm{c}}{R_\mathrm{f}}
.\end{equation}
Current estimates of the relative merger rates of field and cluster mergers remain uncertain and strongly model dependent. Typical literature estimates suggest that field mergers dominate over cluster mergers by approximately one order of magnitude, although the allowed range spans several orders of magnitude \citep{Mandel:2021smh}.
Motivated by these estimates, we adopt 
\begin{equation} \label{eq:Bayesthreshold}
    \log_{10} \bayes \geq 1
\end{equation}
as an heuristic threshold for identifying candidate dense-cluster events. However, different assumptions about the relative merger rates would shift this threshold, motivating one as high as $\log_{10} \bayes \geq 3$ or as low as $\log_{10} \bayes \geq -1$. To account for this uncertainty, Table~\ref{tab:extra} shows the number of candidates selected for different values of the threshold.
}

Events below this threshold are not excluded from having a cluster origin; rather they are not considered robust candidates based on their intrinsic parameters.

\begin{figure*}
    \centering
    \includegraphics[width=1.9\columnwidth]{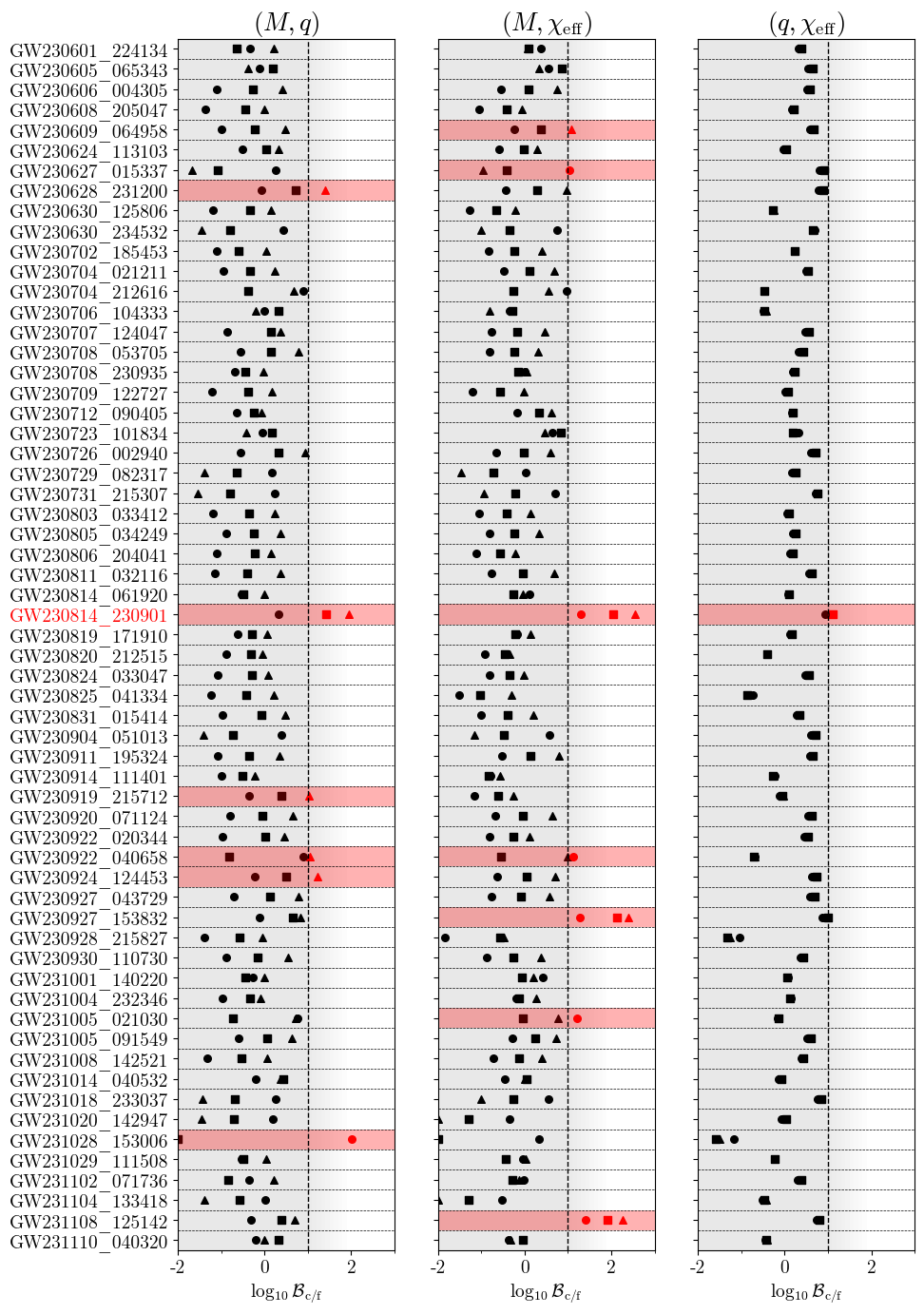}
    \caption{(Continues in the next page)}
    \label{fig:6}
\end{figure*}
\begin{figure*}
    \ContinuedFloat
    \centering
    \includegraphics[width=1.9\columnwidth]{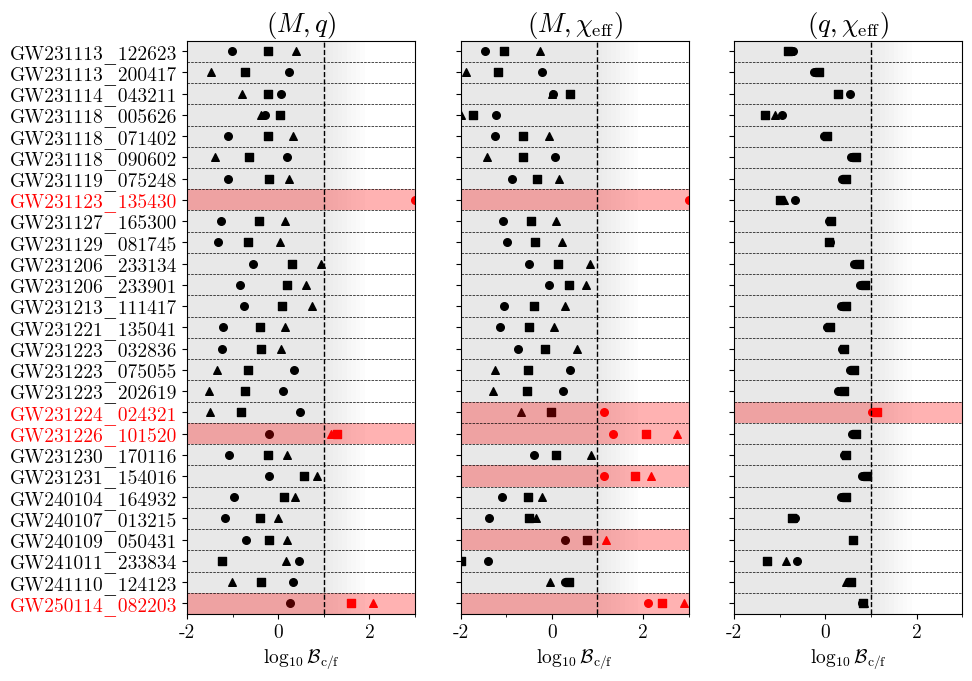}
    \caption{$\log_{10}\bayes$ for all 87 events and the parameters $\mathbf{x}=(M,q)$ (left panel), $\mathbf{x}=(M,\chieff)$ (middle panel), and $\mathbf{x}=(q,\chieff)$ (right panel).
    Events are sorted in chronological order.
    Pairs of populations with metallicities $Z=2\cdot10^{-4}$, $Z=2\cdot10^{-3}$, and $Z=2\cdot10^{-2}$ are indicated by triangle, square and circle markers, respectively. For each event, subspace and cluster metallicity, only the minimum value of the 3 Bayes factors corresponding to the 3 different field populations is shown.
    When a Bayes factor is above the threshold in Eq.~\eqref{eq:Bayesthreshold}, the corresponding point is colored in red. Since it is the minimum of all three Bayes factors corresponding to the three field populations, the column is shaded, indicating that an event is selected.
    Values of $\bayes$ that would otherwise lie outside the plot axis are placed at the top or bottom edges.
    Missing points correspond to vanishing overlap with both cluster and field populations, yielding a $\bayes = \frac{0}{0}$ situation, \revised{but all events have support for at least one of the considered populations.}}
\end{figure*}

\revised{With the procedure described above, we computed Bayes factors for each event, for each choice of parameters $\mathbf{x}$ -- the three possible pairs of $M$, $q$, and $\chieff$ -- and for all nine combinations of cluster and field populations.
Figure~\ref{fig:6} shows $\log_{10}\bayes$ for all events using the three choices of $\mathbf{x}$, one in each column.
For each event, parameter subspace and cluster metallicity, only the smaller Bayes factor among the different field realizations is shown. If this value exceeds the threshold in Eq.~\eqref{eq:Bayesthreshold}, this means that all three Bayes factors for this cluster metallicity exceed the threshold and therefore the cluster-origin hypothesis is preferred over the field-origin. In this case, the point corresponding to this cluster is colored red and the background for this event and parameter subspace is also shaded.
Events displayed in red font are the ones selected as cluster candidates, as explained in Sect.~\eqref{ssec:4.4}.
}

The choice of field prescription can influence the results.
While the prescription chosen for the PSN limit only affects the high-mass tail of the BH mass distribution, the CE development criteria have a much stronger impact on both the total mass and the mass-ratio distributions.
Massive BBHs are more likely to form via stable Roche-lobe overflow (RLOF) evolution since their progenitors are able to avoid stellar merger while entering CE with Hertzsprung gap star donor \citep{Olejak:2021iux}.
The mass-ratio distribution is found to be highly sensitive to the convection growth time parameter, for binaries which evolve through the CE evolution channel as compared to those evolving via stable RLOF \citep{Olejak:2022zee}.

To assess this dependence, we repeated the analysis using field catalogs constructed with the standard CE prescription instead of the revised one, as discussed in Sect.~\ref{ssec:3.2}, and show the results in App.~\ref{app:A}.
Because the standard prescription restricts BH masses more strongly, high-mass events are expected to yield larger positive Bayes factors towards cluster origin under that assumption. 
For this reason, the analysis presented in the main body is more conservative in assigning cluster candidates.

Finally, we emphasize that the Bayes factors defined here are not intended to provide definitive classifications for individual events.
Rather, they establish a quantitative ranking that identifies systems most plausibly associated with dense stellar environments.
In the next section, we use this ranking to select a subset of events for which retention probabilities in dense environments are astrophysically meaningful to evaluate.

\subsection{Cluster candidate selection} \label{ssec:4.4}

The analysis above provides, for each event and parameter pair, a set of nine Bayes factors comparing consistency with cluster and field formation channels.
We now summarize how these results were combined to identify candidate cluster-origin systems.

As discussed in Sect.~\ref{ssec:4.3}, for each subspace $\mathbf{x}$, $(M,q)$, $(M,\chieff)$, or $(q,\chieff)$, we chose those events where all Bayes factors corresponding to a given cluster population are above the threshold defined in Eq.~\eqref{eq:Bayesthreshold}.
These events are highlighted in red in their corresponding panels in Figs.~\ref{fig:6}a.
The number of events selected in at least one of the three analyses is 17.
Of these, 6 events were selected by at least two analyses, and only one (\fullevent{GW230814_23}) by all three analyses.

We considered as \revised{candidate dense-cluster events} those that were selected in at least two subspaces for the same cluster metallicity.
\revised{The requirement that an event exceeds the threshold in at least two parameter subspaces is a pragmatic consistency criterion. Because of the low number of systems of the cluster catalogs, constructing three-dimensional KDEs is challenging, while relying on a single two-dimensional projection can lead to accidental overlaps.
Requiring consistency across multiple complementary subspaces mitigates these projection effects that could drive false positives while remaining computationally robust. For completion, in Table~\ref{tab:extra2} we show the number of events that would have been selected if only 1 or 3 subspaces were required.}

From the 6 events that were selected in at least two subspaces, \fullevent{GW230922_04} was selected with the $Z=2\cdot10^{-4}$ cluster in the $\mathbf{x}=(M,q)$ analysis and with the $Z=2\cdot10^{-2}$ cluster in the $\mathbf{x}=(M,\chieff)$ analysis, and therefore is not on our cluster candidate list.

Then, the list of cluster candidates is \fullevent{GW230814_23}, \fullevent{GW231123},~\fullevent{GW231224},~\fullevent{GW231226}, and \fullevent{GW250114}.
We find that these events come from 3 distinct regions of the parameter space:

\begin{description}
    \item[(1) \emph{Heavy binaries.}]
    \fullevent{GW231123} is the most massive event detected so far by the LVK Collaboration \citep{LIGOScientific:2025rsn,LIGOScientific:2025slb, Xu:2025ajj}.
    This event was selected with the $Z=2\cdot10^{-2}$ cluster population, which corresponds to younger clusters that host more massive binaries.
    In the $\mathbf{x}=(q,\chieff)$ analysis, agnostic to the total binary mass, this event showed $\log_{10}\bayes \sim -0.6$, favoring field origin.
    Other events in this region of the parameter space that are not in our selection of cluster candidates but show mild positive Bayes factors include \fullevent{GW230922_04} and \fullevent{GW231028}.
    \item[(2) \emph{Loud, near-equal mass, negative-$\chieff$ binaries.}] 
    Events \fullevent{GW230814_23}, \fullevent{GW231226}, and \fullevent{GW250114} have $q \sim 1$, masses in the intermediate range and are among the loudest detected so far.
    \fullevent{GW230814_23} has $M\sim 60\ \Msun$ and slight preference for negative $\chieff$ and is the loudest event in \GWTCFOUR with network signal-to-noise ratio of 42 \citep{LIGOScientific:2025cmm, LIGOScientific:2025slb}.
    \fullevent{GW231226} has $M\sim 75\ \Msun$ and $\chieff<0$ at $98\%$ confidence, and constitutes the second loudest \OFOURa event according to \citet{LIGOScientific:2025slb}.
    \fullevent{GW250114} is a special \OFOURb event with $M\sim 65\ \Msun$ and $\chieff<0$ at $98\%$ confidence and a network signal-to-noise ratio of almost 80 \citep{LIGOScientific:2025rid}.
    Such high signal-to-noise ratio cause the posterior samples to occupy a smaller region of the parameter space and therefore to be more susceptible of a sharper preference for a population. It is expected that the loudest events in the catalog have total masses around these values, since systems with such mass emit signals in the frequencies where detectors are the most sensitive.
    All three events were selected for both the $Z=2\cdot10^{-4}$ and $Z=2\cdot10^{-3}$ clusters and show positive Bayes factors in the three analyses.
    Other events in this region of the parameter space that don't show a high enough preference for cluster population include \fullevent{GW230609}, \fullevent{GW230628}, \fullevent{GW230924} and \fullevent{GW230927_15}.
    \item[(3) \emph{Light, similar-mass, zero-$\chieff$ binaries.}] 
    \fullevent{GW231224} is on the lower end of the mass range, with $M \sim 17 \Msun$, similar masses and $\chieff$ is tightly constrained around 0.
    This event was selected for the $Z=2\cdot10^{-2}$ cluster population.
    Other events with $M\lesssim 20\ \Msun$ and $\chieff\sim0$ that are not in the candidate list due to its lower Bayes factors include
    \fullevent{GW230627}.
\end{description}

The parameters of these events highlight regions of parameter space that are more naturally populated by BBHs formed in dense clusters than by isolated field evolution, such as high total masses, near-equal mass ratio, and negative values of $\chieff$.

\begin{figure*}
    \centering
    \includegraphics[width=2\columnwidth]{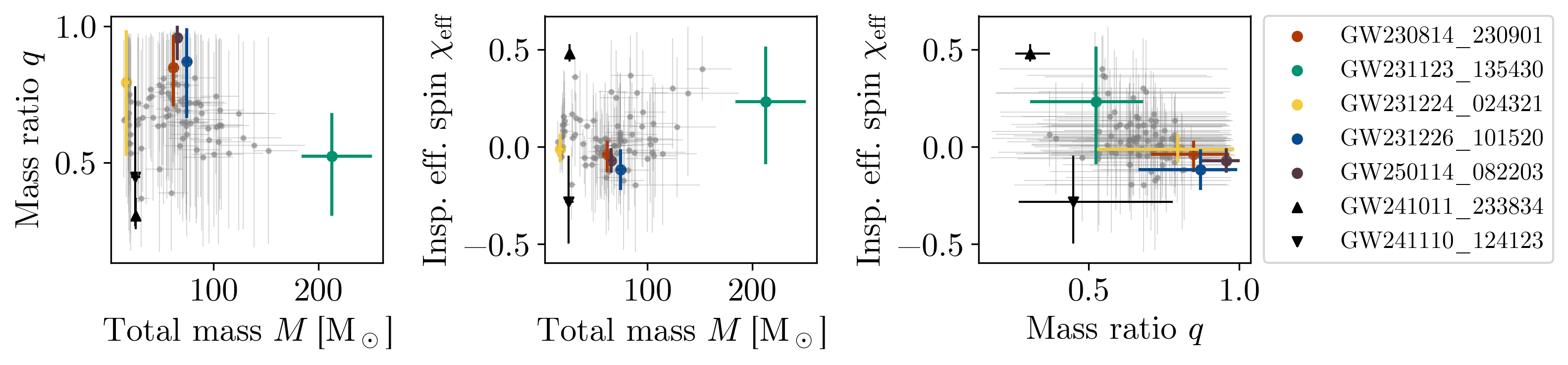}
    \caption{Population distribution of the medians and $90\%$ symmetric credible intervals of the marginalized 1D distributions, for all GW events in the analysis.
    Each panel shows a 2D subspace as indicated in the axes labels.}
    \label{fig:9}
\end{figure*}

Figure~\ref{fig:9} shows the values of the posterior distributions for $M$, $q$, and $\chieff$ of all analyzed events, with the dense-cluster candidate events highlighted in colors.
We can observe how different events lie in specific regions of the parameter space.

In this work, we have assumed that the events come from quasicircular binaries.
This is specifically noted in the use of the quasicircular model \IMRPhenomXPNR for the PE of the GW events.
However, cluster and field binaries are different under this lens: while field binaries become bound in the early inspiral and have had time to circularize before the merger, cluster binaries are usually formed through N-body encounters and might exhibit non-negligible eccentricities at late inspiral phases.

Given that \citet{Xu:2025ajj} presented an analysis of eccentric candidates from \OFOURa using the time-domain aligned-spin eccentric phenomenological waveform model \IMRPhenomTEHM \citep{Planas:2025feq}, it is natural to compare their conclusions with our candidates for cluster origin.
Of the seven events identified in \citet{Xu:2025ajj} as potential eccentric candidates, both \fullevent{GW231123} and \fullevent{GW231224} appear in our final selection of cluster candidates.
However, that study did not ultimately classify \fullevent{GW231123} as eccentric, as Bayes factors favor a precessing interpretation over an eccentric one.
On the contrary, \fullevent{GW231224} shows mild preference for the eccentric hypothesis with $\log_{10}\mathcal{B} = 0.4$ when comparing the eccentric aligned-spin hypothesis with \IMRPhenomTEHM and the quasicircular precessing hypothesis using \IMRPhenomTPHM \citep{Estelles:2021gvs}.
Overall, we observe a modest overlap between the two lists, which warrants further investigation with additional events before drawing any conclusions about potential correlations.
Since eccentricity and precession effects can be difficult to distinguish at low signal-to-noise ratios, reliably incorporating eccentricity in this study would require an eccentric-precessing waveform model that includes multipole asymmetries, which is currently unavailable.

\fullevent{GW241011}~and~\fullevent{GW241110}, detected during \OFOURb, have been independently discussed in the literature as potential products of dense environments.
In particular, \citet{LIGOScientific:2025brd} showed that the component masses and primary spin magnitude of \fullevent{GW241011} are consistent with the ranges predicted for higher-generation mergers in low-metallicity clusters.
Both events are inferred to have large primary spin magnitudes: \fullevent{GW241011} has $\chi_1 \geq 0.7$ at $90\%$ credibility and \fullevent{GW241110} exhibits a posterior peaking around $\chi_1 \sim 0.6$, but with substantial uncertainty. 
However, our analysis differs from that of \citet{LIGOScientific:2025brd} in two main aspects.
First, we did not include individual spin magnitudes, which are generally only weakly constrained in the parameter estimation of real events.
Second, rather than relying on visual overlap with cluster population predictions, we performed a quantitative comparison between cluster and field populations via the computation of Bayes factors.

Neither of these events was flagged as a cluster-origin candidate in our analysis.
As shown in Fig.~\ref{fig:9}, both systems have relatively low total mass (around $\sim 25\ \Msun$) and low mass ratios.
Such low total masses are more naturally present in field formation scenarios.
The most distinctive parameter for these events is $\chieff$. \fullevent{GW241011} has $\chieff \sim 0.5$, while \fullevent{GW241110} lies at $\chieff \leq 0$, though with broad uncertainty.
Although large positive $\chieff$ values are compatible with a cluster origin,
our population models show that they are also prevalent in field binaries, due to the efficient binary spin alignment.
Consequently, our analysis yields Bayes factors that strongly favor a field origin for \fullevent{GW241011} and moderate values for \fullevent{GW241110}.

\revised{As noted in section \ref{ssec:4.3}, the number of selected candidates varies with the choice of threshold in Eq.~\eqref{eq:Bayesthreshold}, which in turn depends on the merger rate ratio. Table~\ref{tab:extra} shows the number of candidates that would be selected for different assumptions.}

\begin{table}
    \centering
    \renewcommand{\arraystretch}{1.2}
    \begin{tabular}{ccc}
    \hline \hline
    $R_\mathrm{c}/R_\mathrm{f}$
    & $\mathcal{B}_\mathrm{c/f}$ threshold & \# of selected candidates \\ \hline
    $10^{-3\phantom{.0}}$   & $\log_{10}\mathcal{B}_\mathrm{c/f} \geq 3$    & 1 (\fullevent{GW231123}) \\
    $10^{-2\phantom{.0}}$   & $\log_{10}\mathcal{B}_\mathrm{c/f} \geq 2$    & 2 (+\fullevent{GW250114}) \\
    $10^{-1.5}$ & $\log_{10}\mathcal{B}_\mathrm{c/f} \geq 1.5$  & 3 (+\fullevent{GW230814_23}) \\
    $\mathbf{10^{-1\phantom{.0}}}$   & $\mathbf{\log_{10}\mathcal{B}_\mathrm{c/f} \geq 1}$    & $\mathbf{5}$ \\
    $10^{-0.5}$ & $\log_{10}\mathcal{B}_\mathrm{c/f} \geq 0.5$  & 31 \\
    $10^{0\phantom{-.0}}$    & $\log_{10}\mathcal{B}_\mathrm{c/f} \geq 0$    & 71 \\
    $10^{1\phantom{-.0}}$    & $\log_{10}\mathcal{B}_\mathrm{c/f} \geq -1$   & 87 \\ \hline
    \end{tabular}
    \caption{\revised{Number of cluster-origin candidate events satisfying posterior odds $\mathcal{O}_\mathrm{c/f}\geq 1$ across a range of assumed merger rate ratios. The 5 events selected for $R_\mathrm{c}/R_\mathrm{f}=0.1$ are the ones derived at the beginning of this section. For higher thresholds, fewer events are selected, and the last column of the table shows which ones are selected for each of the $R_\mathrm{c}/R_\mathrm{f}$ values in the table. Lower thresholds gradually select more events, with the full catalog selected by the threshold $\log_{10}{\mathcal{B}_\mathrm{c/f}} \geq -1$.}}
    \label{tab:extra}
\end{table}

\begin{table}
    \centering
    \renewcommand{\arraystretch}{1.2}
    \begin{tabular}{cc}
    \hline \hline
    \# of subspaces required & \# of selected candidates \\ \hline
    1 & 17 \\
    \textbf{2} & \textbf{5} \\
    3 & 1 (\fullevent{GW230814_23}) \\ \hline
    \end{tabular}
    \caption{\revised{Number of cluster-origin candidate events satisfying posterior odds for a threshold of $\log_{10}\mathcal{B}_\mathrm{c/f} \geq 1$ assuming that a different coincidence is required for selecting events as cluster-origin. In the first row, cluster preference for only 1 2D subspace is enough to select an event; in the second row we show the current choice, while in the third row, where coincidence over all 3 subspaces is required for selecting a candidate, only \fullevent{GW230814_23} is selected.}}
    \label{tab:extra2}
\end{table}

The resulting list of candidate cluster-origin systems forms the basis for the retention analysis in Sect.~\ref{sec:5}. In particular, for these events we evaluated whether their inferred remnant kicks are compatible with retention in globular cluster potentials, thereby assessing the viability of hierarchical merger scenarios in dense environments.

\revised{We emphasize that the present work compares only the adopted globular-cluster population models against selected isolated-binary field populations, and therefore does not exhaustively test all possible dynamical or alternative formation channels, such as AGN disks, young star clusters, or chemically homogeneous evolution.}

\section{Recoil distributions and retention probabilities} \label{sec:5}

In this section we present the recoil velocity distributions inferred for the GW events and assess the retention probability of their merger remnants in different environments.
We first described the kick distributions obtained from \IMRPhenomXPNR posterior samples and then evaluated retention in GCs, NSCs, and galactic potentials.

\subsection{Recoil distributions} \label{ssec:5.1}

For each event, recoil velocities were computed from the \IMRPhenomXPNR waveform using the procedure described in Sect.~\ref{ssec:2.2}.
Rather than evaluating the kick at a single best-fit configuration, we computed it for every posterior sample obtained from parameter estimation, thereby accounting for both observational uncertainty and intrinsic degeneracies in the source parameters.

To quantify the impact of waveform systematics on kicks at the population level, we computed kick distributions for a synthetic set of $N=10^4$ BBH systems drawn from broad priors in mass ratio ($q \in [1/6,1]$), chirp mass ($M_c \in [25,80] \Msun$), and spin magnitudes ($\chi_i \in [0,0.99]$), flat in component masses and with isotropic spin orientations.
For this set, we compared the recoil distributions predicted by several waveform models that include equatorial asymmetries: \IMRPhenomXOFOURa \citep{Thompson:2023ase}, \IMRPhenomXPNR \citep{Hamilton:2025xru}, \SEOBNRvFIVEasym \citep{Estelles:2025zah}, and \NRSurprec \citep{Varma:2019csw}.
Waveforms were generated in the time domain and initialized at a reference frequency $f_\mathrm{ref} = 20 \Hz$.
To transform waveforms from the native frequency-domain models \IMRPhenomXOFOURa and \IMRPhenomXPNR into the time-domain for the kicks computation, the waveforms were transformed as discussed in Sect.~\ref{ssec:2.2}.
The resulting distributions have almost identical median values, which is remarkable given the diversity of modeling approaches.
The symmetric $90\%$ credible intervals are also similar, with \IMRPhenomXOFOURa and \IMRPhenomXPNR extending a little bit further into higher kicks, as reported in Table~\ref{tab:3}.

\begin{table}
    \centering
    \renewcommand{\arraystretch}{1.2}
    \begin{tabular}{cc}
    \hline \hline
    Model & $v_{kick}\ [\kmstight]$ \\ \hline
    \IMRPhenomXOFOURa   & $365^{+1198}_{-273}$ \\
    \IMRPhenomXPNR      & $362^{+1231}_{-266}$ \\
    \SEOBNRvFIVEasym    & $365^{+1096}_{-241}$ \\
    \NRSurprec          & $367^{+1042}_{-248}$ \\ \hline
    \end{tabular}
    \caption{Kick medians and $90\%$ symmetric credible intervals for a synthetic set described in Sect.~\ref{ssec:5.1}, when computed with different waveform models.}
    \label{tab:3}
\end{table}

After validating kick estimates with \IMRPhenomXPNR, 
we computed posterior distributions of recoil velocity for individual BBH events.
For events belonging to \OFOURa, we used the publicly available posterior samples produced by \citet{Xu:2025ajj}.
For the three additional \OFOURb events we performed PE following the \PEAutomator analysis pipeline \citep{PEAutomator} with the same waveform settings to ensure consistency across the sample.

\begin{figure}
    \includegraphics[width=1\linewidth]{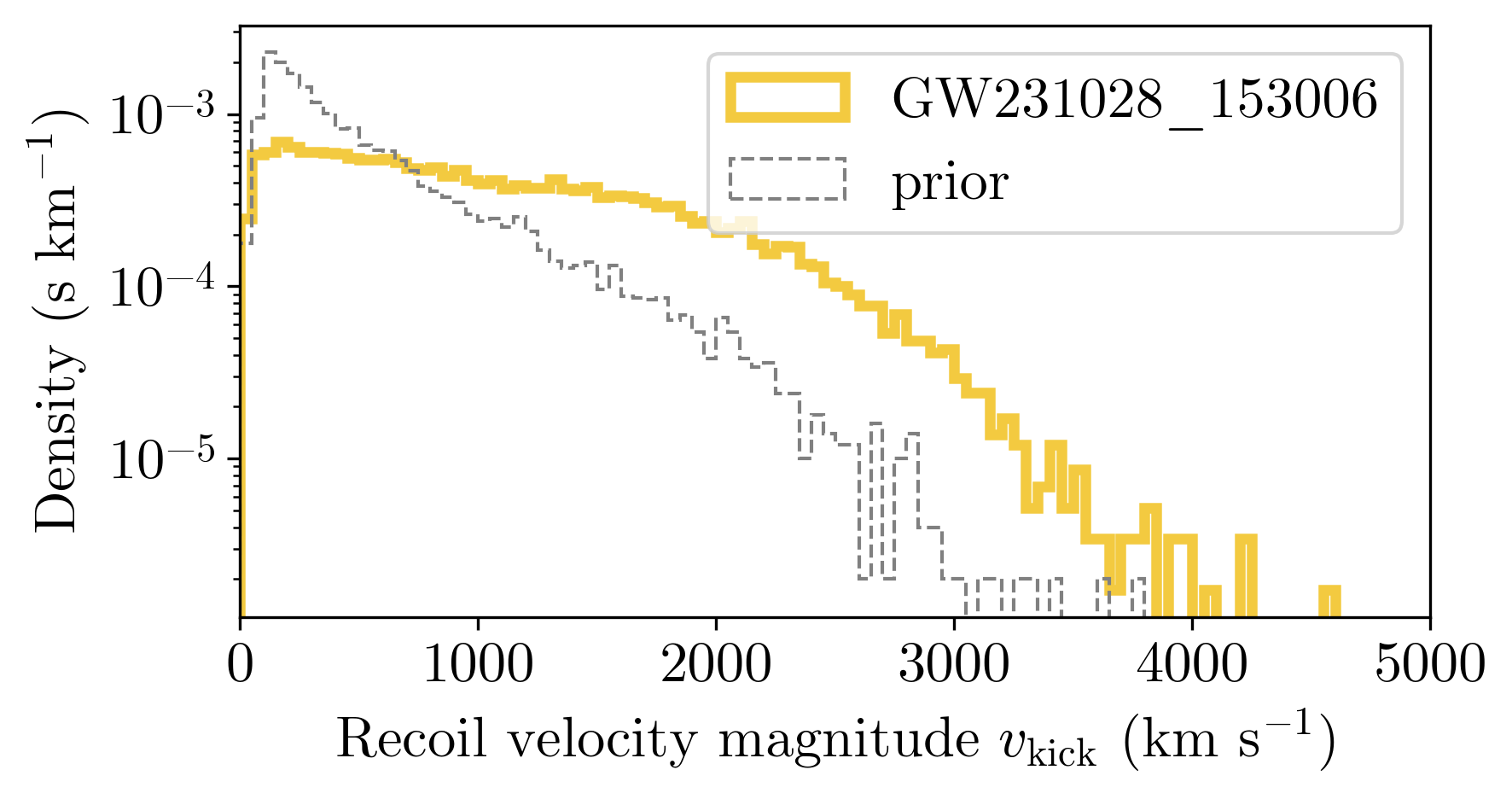}
    \caption{Distribution for the recoil velocity of the \fullevent{GW231028} event, and the prior distribution. We observe the kicks expected from this event are in general larger than the ones in the prior distribution.}
    \label{fig:10}
\end{figure}

For each event, a posterior distribution for the kick magnitude $\vkick$ was computed.
As an illustrative example, Fig.~\ref{fig:10} shows the distribution of the recoil velocity for event \fullevent{GW231028}, compared with that of the prior described at the beginning of the section.

\begin{figure*}
    \includegraphics[width=\columnwidth]{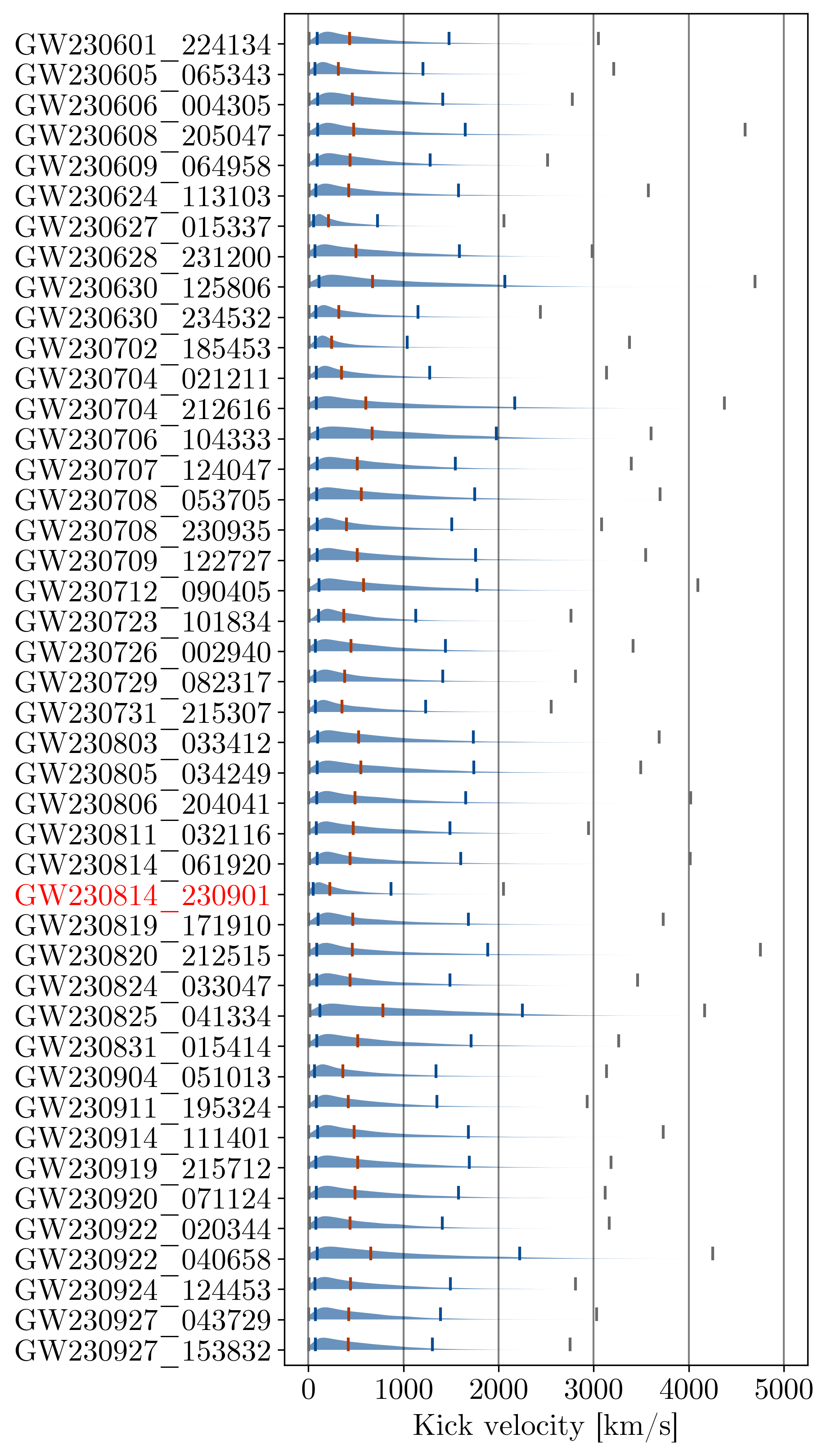}
    \hspace{1em}
    \includegraphics[width=\columnwidth]{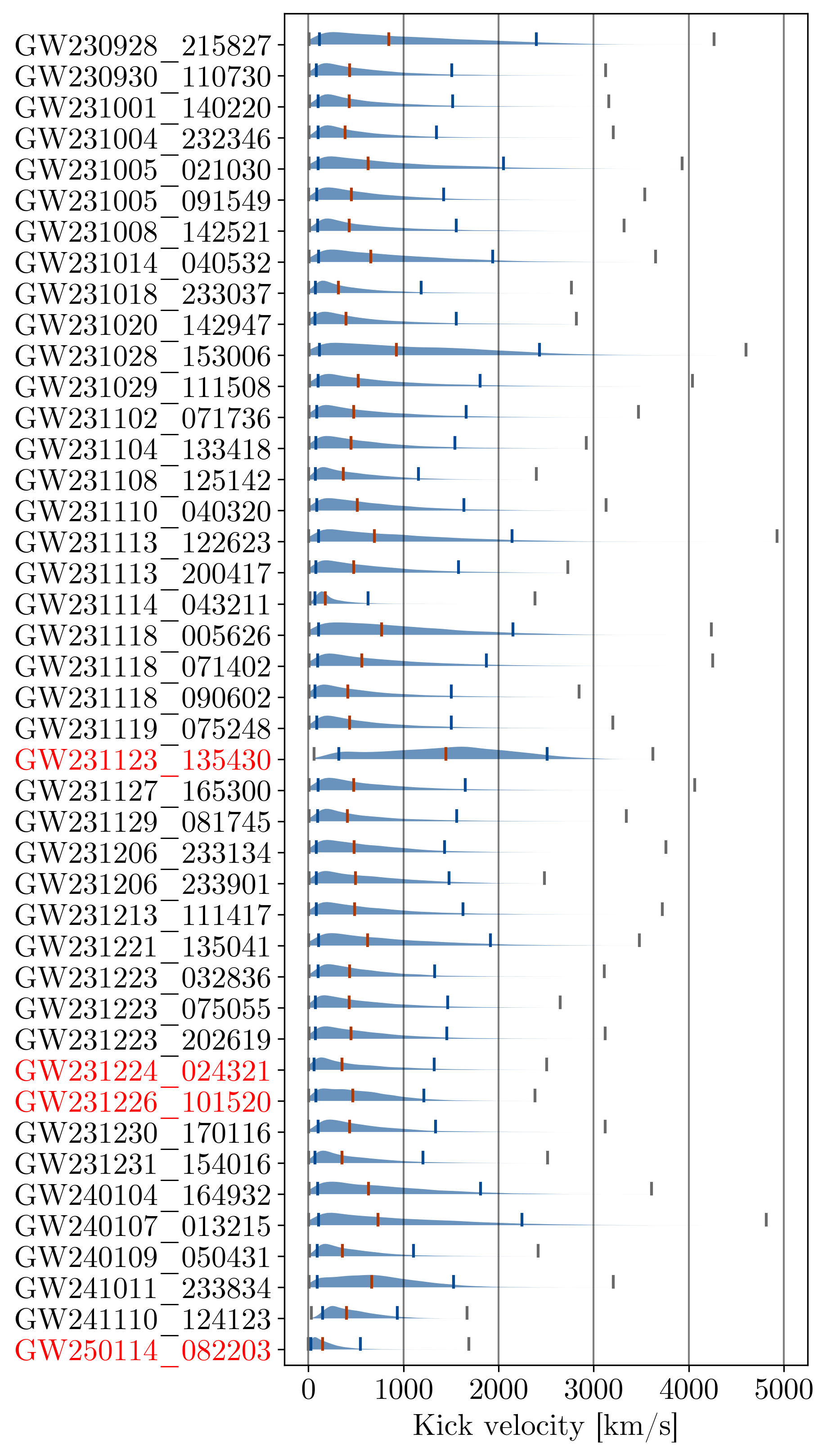}
    \caption{Kick distributions for all analyzed events. On top of the distribution, the red line indicates the median value of the distribution, the blue lines mark the limits of the $90\%$ symmetric credible interval, and the grey marks represent the full range of velocities corresponding to the posterior samples.
    Cluster candidates identified in previous sections are marked in red.}
    \label{fig:11}
\end{figure*}

The kick posterior distributions for all analyzed events are shown in Fig.~\ref{fig:11}.
For the majority of the analyzed BBH mergers, the inferred recoil velocities are modest, with median values around several hundred kilometers per second.
While the majority of events exhibit extended high-velocity tails, very few show large values for the median kick, with the notable exception of \fullevent{GW231123}.

We find that there is no particular correlation between the events identified in Sect.~\ref{ssec:4.4} as cluster candidates and the ones with largest median recoil velocities.
The binary configurations yielding high recoil velocities are not uniquely associated with a specific binary formation channel.
More specifically, binary configurations with high multipole asymmetries like near-equal masses and misaligned high-magnitude spins share some overlap with but are not unique to cluster populations.

In the next section, we compare the kick posteriors with characteristic escape velocities of dense stellar environments to quantify the probability that merger remnants are retained within, or ejected from, their host systems.

\subsection{Retention probability} \label{ssec:5.2}

The astrophysical impact of recoil velocities depends on whether the merger remnant remains gravitationally bound to its host environment.
Given the posterior distribution of recoil velocities for an event, we quantified this by computing the probability that the remnant is retained within a dense stellar system.

For an event with kick-velocity probability distribution $f_\mathrm{event}(v)$, and a characteristic escape velocity $\vesc$ of the host environment, the retention probability is defined as
\begin{equation}
    \mathcal{R}(\mathrm{event}) = \int_0^{\vesc} f_\mathrm{event}(v)\ \mathrm{d}v \cong \dfrac{1}{N_\mathrm{samp}} \sum_{i=1}^{N_\mathrm{samp}} \mathbb{1}_{v_i \leq \vesc}
,\end{equation}
where $\{ v_i \}_{i=1}^{N_\mathrm{samp}}$ are the kick velocities computed from the PE posterior samples for the event,
while the corresponding ejection probability is 
\begin{equation}
    \mathcal{E}(\mathrm{event}) = 1 - \mathcal{R}(\mathrm{event})
.\end{equation}
These quantities represent the fraction of posterior support for which the remnant black hole remains bound to, or escapes from, the host potential, respectively.

The escape velocity depends strongly on the nature of the host environment.
Based on present-day observed properties of Galactic GCs, their central escape velocities are typically of order $\sim 50-100 \kms$ \citep{2002ApJ...568L..23G,2018MNRAS.478.1520B,2020PASA...37...46B}.
Owing to their much higher central densities and different formation histories, nuclear star clusters can reach central escape velocities of several hundred kilometers per second \citep{Antonini:2016gqe,PhysRevD.100.041301,2019MNRAS.486.5008A,2020MNRAS.498.4591F}.
\revised{We evaluate retention in NSCs independently of the fact that retention was identified using GC population models, and no conclusions should be drawn on the possible NSC-origin of such events.}

Here, we adopted escape velocities toward the upper end of these ranges, namely $\vesc = 100 \kms$ for GCs and $\vesc = 600 \kms$ for NSCs.
These choices intentionally bias retention probabilities toward larger values, thereby providing conservative estimates of remnant ejection.
Since the bulk of the kick distribution lies around these values, the derived retention probabilities are sensitive to the assumed threshold velocity; so adopting higher values ensures that our conclusions do not overestimate ejection rates.

\begin{figure}
    \includegraphics[width=\columnwidth]{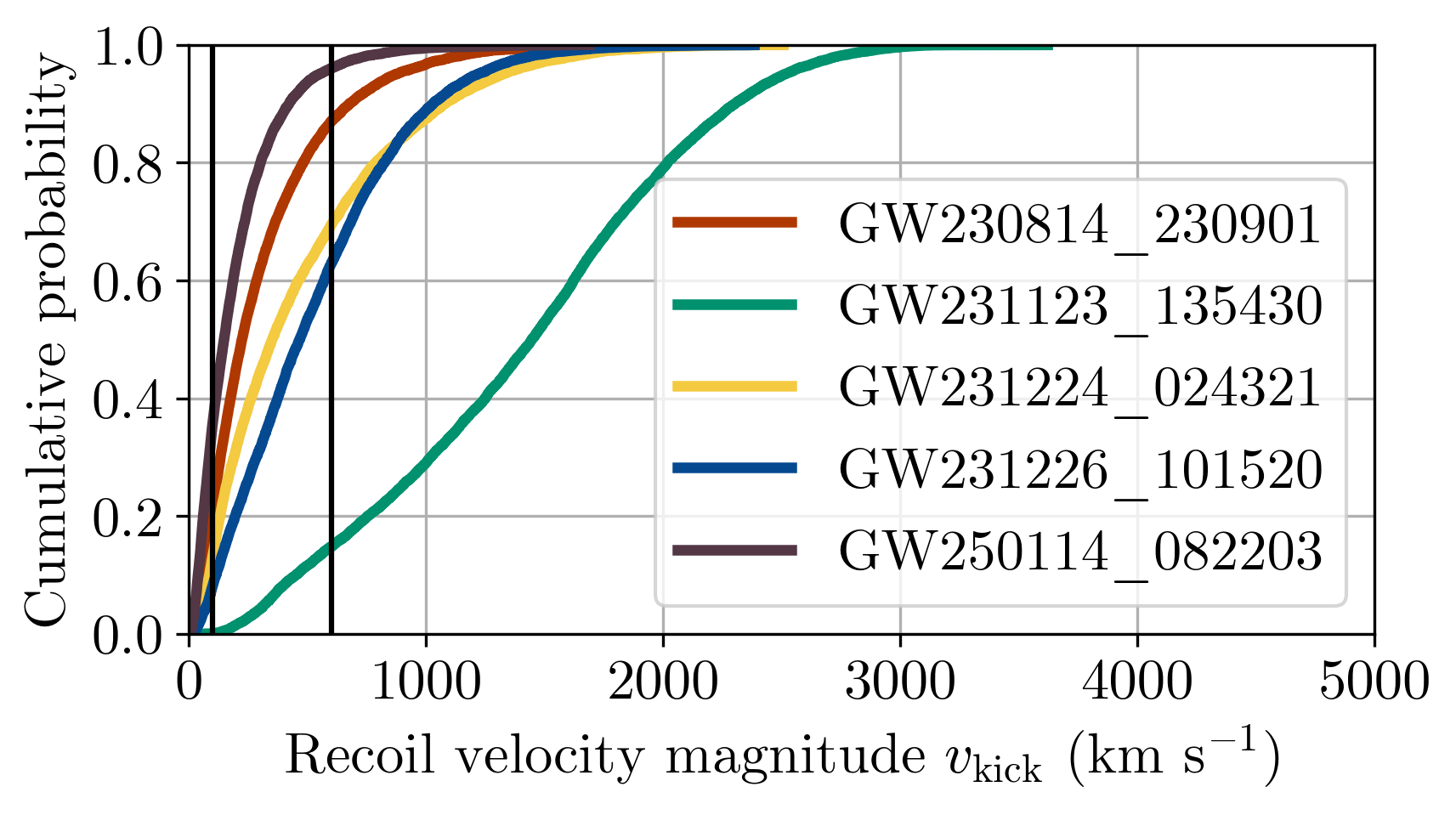}
    \caption{Cumulative distribution functions for the kick magnitude of the five cluster-origin candidate events, with black lines marking the assumed $\vesc$ thresholds for GCs ($100\kms$) and NSCs ($600\kms$).}
    \label{fig:12}
\end{figure}

We computed retention probabilities for the subset of events identified in Sect.~\ref{ssec:4.4} as likely cluster-origin candidates, where these systems formed and merged within dense stellar environments.
The resulting ejection probabilities for GCs and NSCs are reported in Table~\ref{tab:4}.
For most events, the inferred recoil distributions imply a high probability of ejection from GCs, with only a small fraction of posterior support lying below the adopted escape-velocity threshold.
In contrast, results for NSCs are more diverse, as their escape velocities are more comparable to the typical recoil magnitudes.
Figure~\ref{fig:12} shows the kick distributions for the five cluster-candidate events, along with vertical lines marking the thresholds $v_\mathrm{GC}=100\kms$, $v_\mathrm{NSC}=600\kms$.
We found that \fullevent{GW231123}, \fullevent{GW231224}, and \fullevent{GW231226} escape a typical GC potential with $\geq90\%$ probability.
In NSC environments, \fullevent{GW250114} is the only event retained with 90\% certainty, while none are ejected at this level of certainty.

\begin{table}
    \centering
    \begin{tabular}{cccc}
    \hline \hline
    & Event & $\mathcal{E}_\mathrm{GC}$ & $\mathcal{E}_\mathrm{NSC}$ \\ \hline    
    \cir{fill=GW230814}{draw=white} & \fullevent{GW230814_23} & $79\%$ & $13\%$ \\  
    \cir{fill=GW231123}{draw=white} & \fullevent{GW231123}    & $99.94\%$ & $85\%$ \\
    \cir{fill=GW231224}{draw=white} & \fullevent{GW231224}    & $95\%$ & $52\%$ \\    
    \cir{fill=GW231226}{draw=white} & \fullevent{GW231226}    & $92\%$ & $37\%$ \\    
    \cir{fill=GW250114}{draw=white} & \fullevent{GW250114}    & $64\%$ & $4\%$ \\ 
    \hline
    \end{tabular}
    \caption{Ejection probabilities for globular clusters (GC) and nuclear star clusters (NSC) for the 5 cluster candidate events reported in Sect.~\ref{ssec:4.4}.}
    \label{tab:4}
\end{table}

We also considered a more general scenario in which BBH mergers occur within the gravitational potential of a massive galaxy, independent of whether the progenitor binary formed in a dense cluster.
We adopted a characteristic escape velocity of $\vesc = 2500 \kms$ \citep{Merritt:2004xa}, representative of giant elliptical galaxies, while noting that the local escape velocity of the Milky Way near the Sun’s position is approximately $\vesc = 600 \kms$ \citep{2018A&A...616L...9M}. 

Under this assumption, we found that the probability that at least one remnant among the 87 analyzed events escapes the galactic potential is non-negligible.
Specifically, assuming that the escape velocity of each event follows the distribution shown above and that they are mutually independent, we estimated the probabilities for at least one, two, and three merger-remnant ejections to be $38\%$, $8\%$, and $1.2\%$, respectively.

These values highlight that even in deep gravitational potentials, extreme recoil events remain astrophysically plausible, and ejected remnants cannot be excluded.

We emphasize that retention probabilities depend sensitively on both the recoil distributions and the assumed escape velocities, and should therefore be interpreted as conditional on the adopted environmental model.
In particular, one could compute the kick distribution using PE posteriors obtained with astrophysical priors corresponding to GC populations, rather than the agnostic priors adopted in this work. This approach is not pursued here because it would require selecting a specific cluster metallicity.

\section{Conclusions}

In this work, we have examined the recoil distributions of BBH mergers observed during \OFOURa and three selected \OFOURb events, along with their astrophysical implications.
We combined the analysis of synthetic BBH population catalogs with parameter estimation of the GW detections in order to identify which events are most likely to come from cluster environments.
We took advantage of recent developments of waveform modeling of multipole asymmetries to estimate the remnant kicks of the various BBH mergers and combine these with observational estimates of escape velocities in dense astrophysical environments to estimate retention probabilities of the final black holes, which influences its ability to undergo hierarchical mergers.

Our results indicate that the majority of events are consistent with formation through isolated binary evolution, which is expected to dominate in the Universe, while we identify 5 events that show statistical preference for \revised{dense-cluster origin}.

These results are obtained looking at how the event PE posterior distributions for parameters $M$, $q$, $\chieff$ compare with the distributions expected by synthetic population catalogs.
We have used state-of-the-art GC catalogs accounting for varying metallicities and formation epoch, and field population catalogs with a revised mass-transfer stability framework.
As seen in Appendix~\ref{app:A}, using the standard prescription for the latter, we find more than two thirds of all events show preference for \revised{GC-like dense-environment origin}, driven by the lower BH mass achieved using this prescription.

These results highlight the importance of quantitative population comparisons, which can yield interpretations that differ from expectations based solely on individual source parameters.
Specifically, our analysis does not find enough evidence to consider \fullevent{GW241011} a cluster-origin candidate due to its low masses and high $\chieff$ being more consistent with the field population than with the cluster.
When compared with potential eccentric candidates, the cluster-origin candidates show moderate overlap, emphasizing the necessity for generic waveform models capable of incorporating orbital eccentricity in future analyses.
While eccentricity is an expected signature of dynamical formation, only a fraction of dynamically produced BBHs show a measurable eccentricity \citep{Rodriguez:2018pss, Samsing:2017oij, Zevin:2018kzq} and high eccentricities are also present in BBH mergers that coming from field binaries \citep{Antonini:2017ash}.
Future analyses could include eccentricity as one of the studied parameters, using an eccentric precessing waveform model to compute the kicks.

We computed the recoil from the gravitational waveform for each of our observed BBH mergers.
The inferred recoil velocity distributions of detected BBH merger remnants are typically of order a few hundred$\kms$, with broad posteriors.
When interpreted in the context of astrophysical environments, we found that most merger remnants associated with globular cluster environments are likely to be ejected, whereas retention becomes more plausible in nuclear star clusters.
Specifically, of the five events found by our analysis to favor \revised{dense-cluster origin}, we found that three are expected to escape a typical GC potential at 90\% confidence, while at least one would be retained by a NSC at this level of certainty.

Although recoil velocities in the order of thousands of$\kms$ remain uncommon for the observed BBH mergers, the probability that at least one remnant exceeds escape velocity from the most massive galactic potentials is non-negligible.
This points towards efficient ejection of some merger remnants and supports the possibility that GW observations begin to probe the production of intergalactic black holes.

The conclusions of this study are nevertheless subject to several sources of uncertainty.
Environmental classification depends on the adopted population models representing different formation channels, as well as on our incomplete knowledge of their relative abundances in the local Universe.
\revised{Our conclusions are conditional on the adopted population models and the approximate detectability treatment.
This analysis does not include other possible BBH formation channels such as AGN disks, young star clusters, or other isolated-binary prescriptions.}
\revised{The resulting Bayes factors should therefore be interpreted primarily as a model-dependent ranking statistic rather than a definitive classification of formation channel.}
Specifically, we found our ability to consider larger-dimensional parameter distributions limited by the small population in GC catalogs.
Since assumptions on the cluster initial properties can affect the parameter distributions \citep{Hong:2018bqs}, a future analysis would benefit from a larger set of cluster models with different initial binary fractions and consistent treatment for binary evolution processes like mass-transfer and common envelope evolution.
The development of expanded and more realistic astrophysical population models will enable a more precise assessment of BBH formation pathways.
On the observational side, events with lower signal-to-noise ratio yield broad posterior distributions and parameter degeneracies, which complicates robust classification.
Future developments in GW detectors will increase their sensitivity and enable higher-quality detections with tighter parameter constraints.

Overall, this work demonstrates that current GW observations already provide meaningful constraints on both the formation channels of BBH sources and the interaction of merger remnants with their host environments.
As observational samples grow and population models improve, GW detections will continue to refine our understanding of the evolutionary channels of BBHs.

\begin{acknowledgements}
The authors would like to thank Filippo Santoliquido for the LSC Publication \& Presentation Committee Review of this manuscript and Sharan Banagiri, Thomas Dent and Juan Calderón-Bustillo for their useful comments.

This work was supported by the Universitat de les Illes Balears (UIB) with funds from the Programa de Foment de la Recerca i la Innovació de la UIB 2024-2026 (supported by the yearly plan of the Tourist Stay Tax ITS2023-086);
the Spanish Agencia Estatal de Investigación grants PID2022-138626NB-I00, RED2024-153978-E, RED2024-153735-E, funded by MICIU/AEI/10.13039/501100011033 and the ERDF/EU;
and the Comunitat Autònoma de les Illes Balears through the Conselleria d’Educació i Universitats with funds from the European Regional Development Fund (SINCO2022/18146 - Plataforma HiTech-IAC3-BIO).

JLQ and FARV were supported through the Conselleria d’Educació i Universitats del Govern de les Illes Balears in the framework of the Balearic Islands ESF+ Program 2021-2027 via FPI-CAIB doctoral grants FPI\_093\_2022, FPI\_092\_2022. 
EH, NS and MC were financed by the Conselleria d'Educaci\'{o} i Universitats del Govern de les Illes Balears and the European Social Fund Plus (ESF+) 2021-2027, through grants POSTDOC2024\_25, POSTDOC2024\_55 and POSTDOC2024\_52. NS is also supported by Programa ``Beques Santander - Investigació Postdoctoral''.
AA acknowledges support for this research from the Polish National Science Center (NCN) grant number 2024/55/D/ST9/02585.
TB was supported by the NCN grant 2023/49/B/ST9/02777.
AO is partially supported by a grant from the German-Israeli Foundation for Scientific Research and Development (GIF; Grant No. I-873-303.5-2024).
SH was also partly supported by the Spanish program Unidad de Excelencia María de Maeztu CEX2020-001058-M, financed by MCIN/AEI/10.13039/501100011033, and by the MaX-CSIC Excellence Award MaX4-SOMMA-ICE.
YX was supported by the INVESTIGA@UIB programme of the Universitat de les Illes Balears (UIB), co-funded by the 2023 Sustainable Tourism Promotion Plan (ITS2023-086 – Research Promotion Programme).  
JV was supported by the Spanish Ministerio de Ciencia, Innovación y Universidades Grant No. FPU22/02211.

We thankfully acknowledge the computer resources at MareNostrum5 Supercomputer, technical expertise and assistance provided by the Barcelona Supercomputing Center (BSC) through funding from the Red Española de Supercomputación (RES) grant number AECT-2025-2-0045;
and the computer resources at Picasso Supercomputer, the technical support and assistance provided by the Supercomputing and Bioinnovation Center (SCBI) of the University of Málaga through funding from grants AECT-2025-2-0019, AECT-2025-2-0025, and AECT-2025-2-0029.

This research has made use of data or software obtained from the Gravitational Wave Open Science Center (gwosc.org), a service of the LIGO Scientific Collaboration, the Virgo Collaboration, and KAGRA. This material is based upon work supported by NSF's LIGO Laboratory which is a major facility fully funded by the National Science Foundation, as well as the Science and Technology Facilities Council (STFC) of the United Kingdom, the Max-Planck-Society (MPS), and the State of Niedersachsen/Germany for support of the construction of Advanced LIGO and construction and operation of the GEO600 detector. Additional support for Advanced LIGO was provided by the Australian Research Council. Virgo is funded, through the European Gravitational Observatory (EGO), by the French Centre National de Recherche Scientifique (CNRS), the Italian Istituto Nazionale di Fisica Nucleare (INFN) and the Dutch Nikhef, with contributions by institutions from Belgium, Germany, Greece, Hungary, Ireland, Japan, Monaco, Poland, Portugal, Spain. KAGRA is supported by Ministry of Education, Culture, Sports, Science and Technology (MEXT), Japan Society for the Promotion of Science (JSPS) in Japan; National Research Foundation (NRF) and Ministry of Science and ICT (MSIT) in Korea; Academia Sinica (AS) and National Science and Technology Council (NSTC) in Taiwan.
\end{acknowledgements}

\bibliographystyle{bibtex/aa.bst}
\bibliography{bibliography}

\begin{appendix}

\section{Results with the standard CE development prescription for field binaries} \label{app:A}

\begin{figure}[h!]
    \begin{subfigure}{\columnwidth}
        \includegraphics[width=0.9\columnwidth]{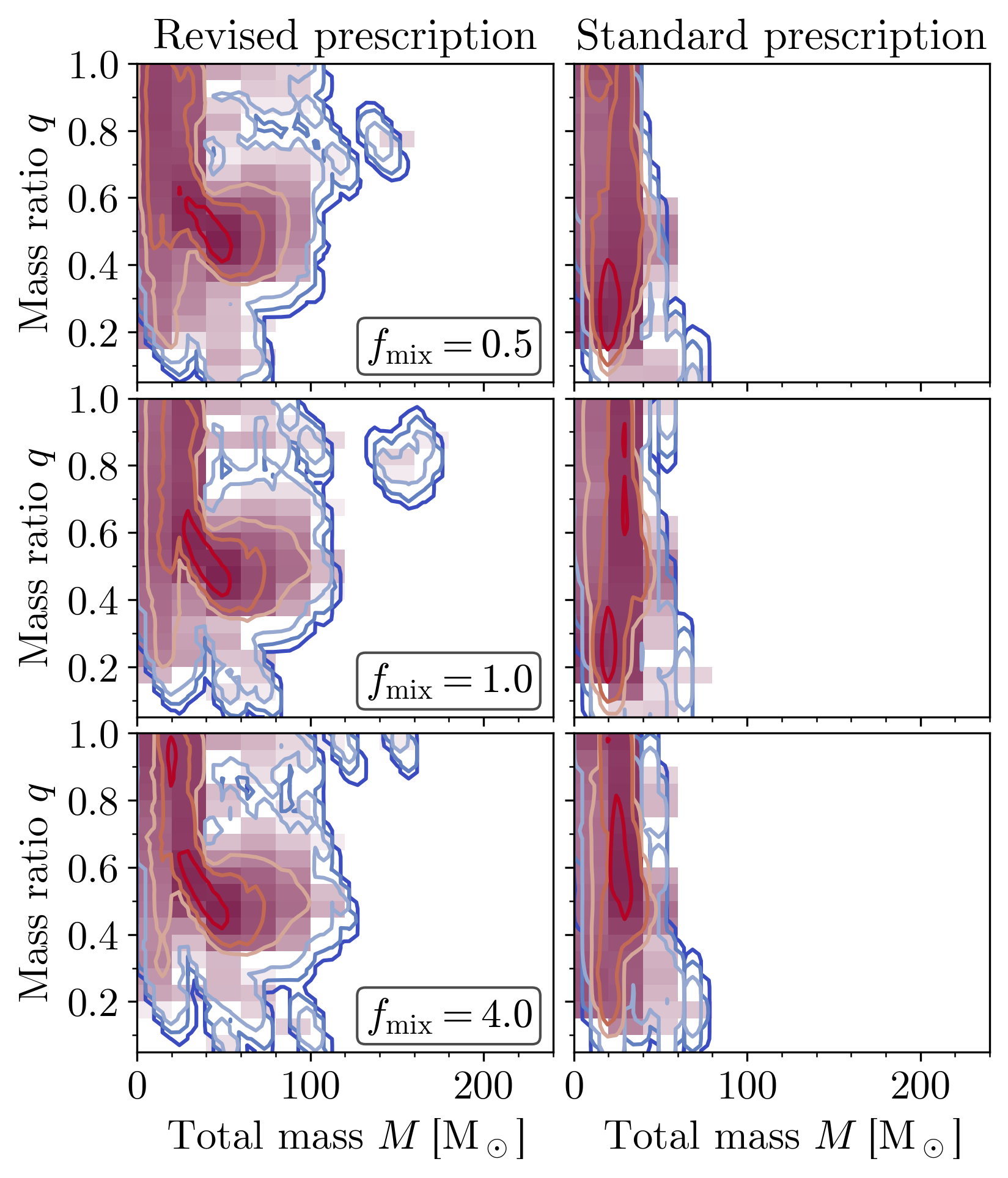}
    \end{subfigure}
    \begin{subfigure}{\columnwidth}
        \centering
        \includegraphics[width=0.5\linewidth]{fig_4d.png}
        \includegraphics[width=0.45\linewidth]{fig_4e.png}
    \end{subfigure}
    \caption{Population distribution for $\mathbf{x}=(M,q)$, both as two-dimensional histograms and KDEs.
    Left column shows distributions with the revised prescription (and is therefore equivalent to the right column of Fig.~\ref{subfig:4a}) and right column shows the equivalent with the standard prescriptions.
    Values of the mixing parameter $f_\mathrm{mix}$ apply to both panels in the same row.
    }
    \label{fig:A1}
\end{figure}

As introduced in Sect.~\ref{ssec:3.2}, field population catalogs adopt different prescriptions that can significantly affect their parameter distributions.
In particular, the CE phase development criteria can strongly influence the mass distributions.
Figure~\ref{fig:A1} shows the mass distributions obtained with the revised and standard prescriptions, where the crucial difference arises from the limitation of the total binary mass imposed by the standard prescription.

Results presented in this paper adopt revised mass-transfer criteria for the RLOF \citep{2017MNRAS.465.2092P,2021A&A...651A.100O}, under which the formation of massive BBHs is more likely.
In this Appendix we present the same analysis using the more standard CE phase development criteria of \citet{2008ApJS..174..223B}.

\begin{figure*}
    \centering
    \includegraphics[width=1.9\columnwidth]{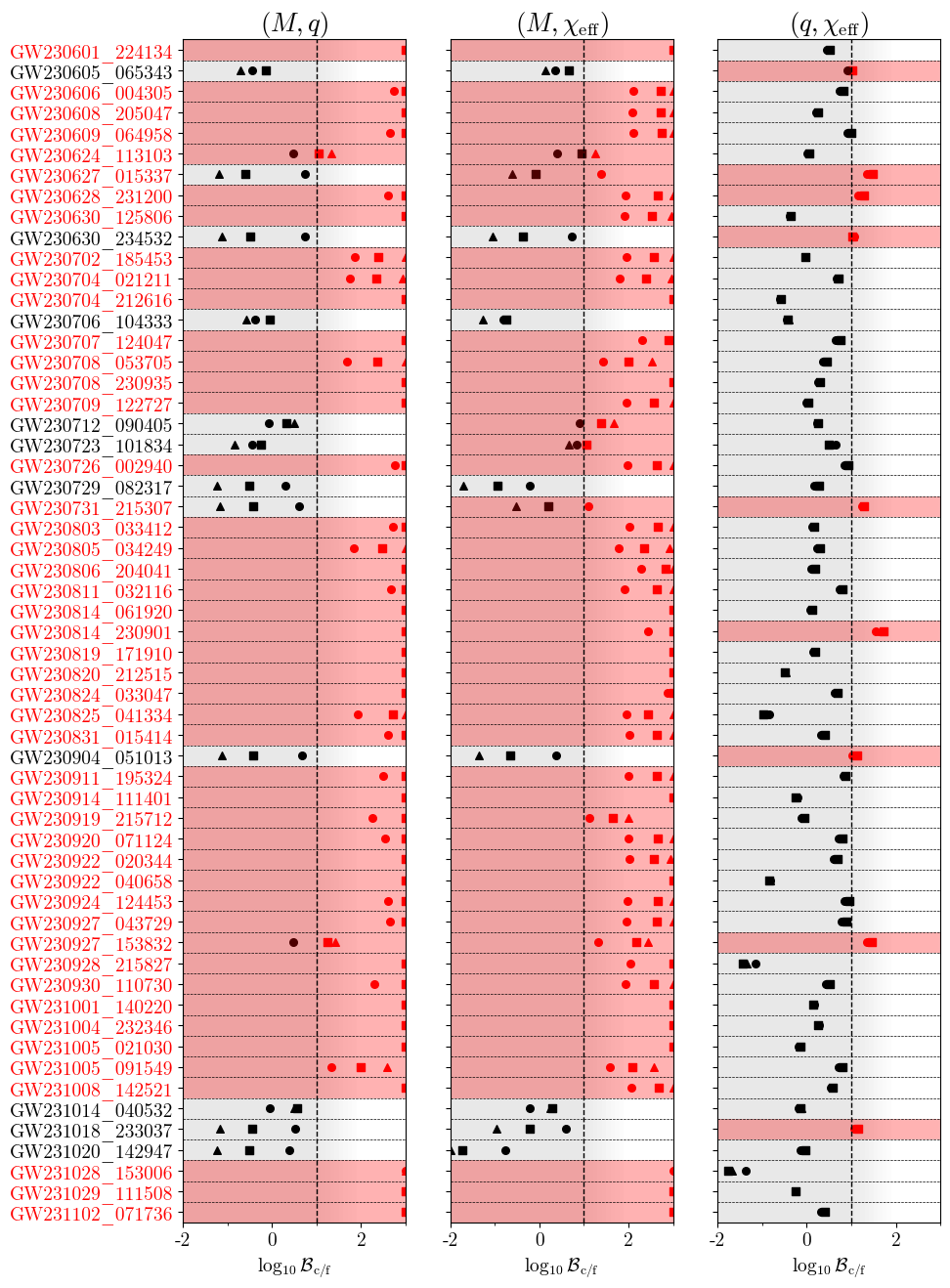}
    \caption{(Continues in the next page)}
\end{figure*}
\begin{figure*}
    \ContinuedFloat
    \centering
    \includegraphics[width=1.9\columnwidth]{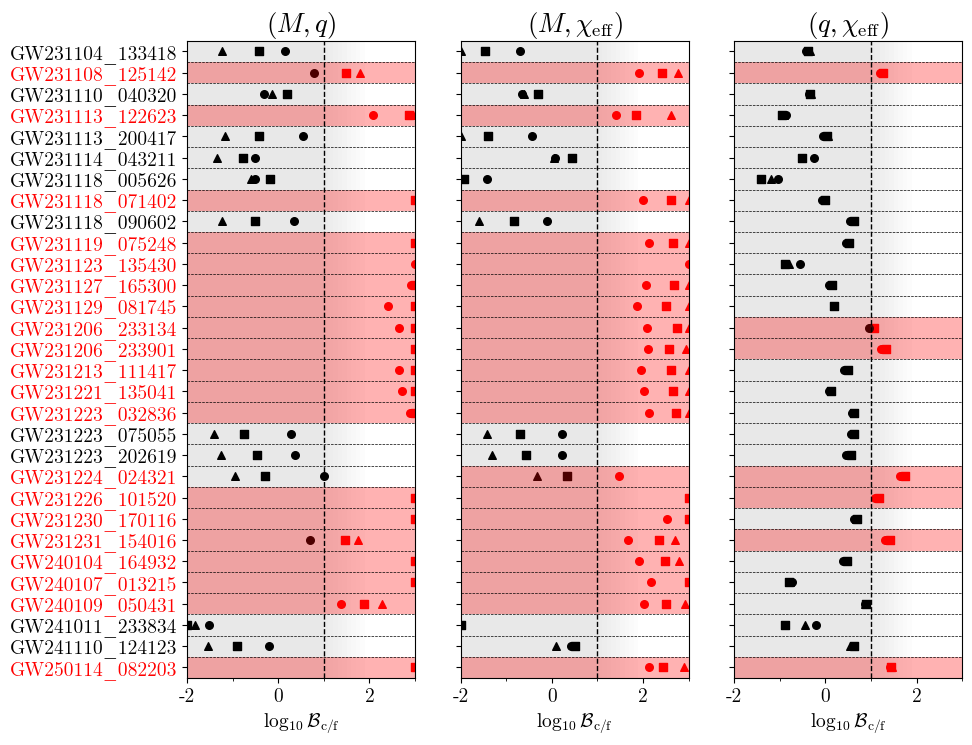}
    \caption{$\log_{10}\bayes$ for all 87 events and the parameters $\mathbf{x}=(M,q)$ (left panel), $\mathbf{x}=(M,\chieff)$ (middle panel), and $\mathbf{x}=(q,\chieff)$ (right panel), using the standard CE prescription.
    Equivalent of Fig.~\ref{fig:6}. 
    Events are sorted in chronological order.
    Pairs of populations with metallicities $Z=2\cdot10^{-4}$, $Z=2\cdot10^{-3}$, and $Z=2\cdot10^{-2}$ are indicated by triangle, square and circle markers, respectively. For each event, subspace and cluster metallicity, only the minimum value of the 3 Bayes factors corresponding to the 3 different field populations is shown.
    When a Bayes factor is above the threshold in Eq.~\eqref{eq:Bayesthreshold}, the corresponding point is colored in red. Since it is the minimum of all three Bayes factors corresponding to the three field populations, the column is shaded, indicating that an event is selected.
    Values of $\bayes$ that would otherwise lie outside the plot axis are placed at the top or bottom edges.
    Missing points correspond to vanishing overlap with both cluster and field populations, yielding a $\bayes = {0}/{0}$ situation; \revised{however, all events in this situation have non-negligible support for at least one cluster population.}}
    \label{fig:A2}
\end{figure*}

\revised{Figure~\ref{fig:A2} shows the Bayes factors computed using the standard prescription.
A substantial increase in the number of selected events is observed, now exceeding two thirds of the catalog. This can be directly attributed to the very limited support that field models with the standard prescription provide for binaries with $M \geq 60 \Msun$, which causes a lot of events being selected for both $\mathbf{x}=(M,q)$ and $\mathbf{x}=(M,\chieff)$.
Several events show no overlap with the field populations, while still overlapping with cluster populations, which results in $\log_{10}\bayes=+\infty$ for a considerable number of cases.
The subspace with $\mathbf{x}=(q,\chieff)$ does not include the total mass $M$ and is therefore less affected by the change of prescription. Nevertheless, 14 additional events are selected, on top of the only 2 selected by the analysis using the revised prescription.}

With this prescription, 67 events would be identified as cluster-origin candidates.
This demonstrates how drastically the assumed field population affects our ability to consider parameters other than the total mass to identify cluster-origin candidates.
If the threshold in Eq.~\eqref{eq:Bayesthreshold} were raised to $+\infty$, so that only events with no possible support from field binaries were considered, 9 events would still be selected, corresponding to the most massive events in \OFOURa.

This result is consistent with literature since the stable mass transfer channel tends to produce patterns in $\mathbf{x}=(M,\chieff)$~and~$\mathbf{x}=(q,\chieff)$ that are similar to what is inferred from GW detections.
This is a consequence of mass-ratio reversal being a common outcome in this channel, followed by tidal spin-up of the secondary BH \citep{Olejak:2024qxr,Banerjee:2024wbq}.
In contrast, the standard CE channel seems to produce trends that are somewhat opposite to what is observed, in addition to lower total masses compared to the revised CE scenario.

\end{appendix}

\end{document}

%% file: commands.tex
\usepackage{xspace}

\newcommand{\Lhat}{\mathbf{\hat{L}}}

\newcommand{\chieff}{\chi_\mathrm{eff}}
\newcommand{\chip}{\chi_\mathrm{p}}

\newcommand{\chii}{\boldsymbol{\chi}_i}

\newcommand{\chiipar}{\chi_i^\parallel}
\newcommand{\chionepar}{\chi_1^\parallel}
\newcommand{\chitwopar}{\chi_2^\parallel}

\newcommand{\chiiperp}{\chi_i^\perp}
\newcommand{\chioneperp}{\chi_1^\perp}
\newcommand{\chitwoperp}{\chi_2^\perp}

\newcommand{\vesc}{v_\mathrm{esc}}
\newcommand{\vkick}{v_\mathrm{kick}}

\usepackage{xcolor}
\definecolor{dodgerblue}{HTML}{1E90FF}
\definecolor{viennared}{HTML}{DA0A14}
\definecolor{ctorange}{HTML}{FF6C0C}
\definecolor{granadagreen}{HTML}{078931}
\definecolor{wales}{HTML}{ff0038}
\definecolor{valenciacfred}{HTML}{ee3524}
\definecolor{barcelonafcgold}{HTML}{edbb00}
\definecolor{jam}{HTML}{A50B5E}
\definecolor{austriawien}{HTML}{441678}
\definecolor{navyblue}{HTML}{4169E1}

\definecolor{DraftColor}{RGB}{30,45,100}

\newcommand{\soft}[1]{\textsc{#1}}
\newcommand{\GWTCFOUR}{GWTC-4.0\xspace}

\newcommand{\astropy}{\soft{Astropy}\xspace}
\newcommand{\bilby}{\soft{Bilby}\xspace}
\newcommand{\dynesty}{\soft{Dynesty}\xspace}
\newcommand{\PEAutomator}{\soft{PEAutomator}\xspace}

\newcommand{\gwfish}{\soft{gwfish}\xspace}

\newcommand{\IMRPhenomXHM}{\soft{IMRPhenomXHM}\xspace}
\newcommand{\IMRPhenomXPHM}{\soft{IMRPhenomXPHM}\xspace}
\newcommand{\IMRPhenomXOFOURa}{\soft{IMRPhenomXO4a}\xspace}
\newcommand{\IMRPhenomXPNR}{\soft{IMRPhenomXPNR}\xspace}
\newcommand{\IMRPhenomTPHM}{\soft{IMRPhenomTPHM}\xspace}
\newcommand{\IMRPhenomTEHM}{\soft{IMRPhenomTEHM}\xspace}
\newcommand{\SEOBNRvFIVEasym}{\soft{SEOBNRv5PHM\_asym}\xspace}

\newcommand{\NRSurprec}{\textsc{NRSur7dq4}\xspace}

\newcommand{\NRSurRemnant}{\textsc{NRSurRemnant}\xspace}

\newcommand{\OTHREE}{\soft{O3}\xspace}
\newcommand{\OFOURa}{\soft{O4}a\xspace}
\newcommand{\OFOURb}{\soft{O4}b\xspace}

\newcommand{\kms}{\ \mathrm{km}\ \mathrm{s}^{-1}}
\newcommand{\kmstight}{\mathrm{km}\ \mathrm{s}^{-1}}
\newcommand{\Gyr}{\ \mathrm{Gyr}}
\newcommand{\Hz}{\ \mathrm{Hz}}
\newcommand{\Msun}{\ \mathrm{M}_\odot}
\newcommand{\ms}{\ \mathrm{ms}}

\newcommand{\event}[1]{\soft{#1}}
\newcommand{\bayes}{\mathcal{B}_\mathrm{c/f}}

\newcommand{\UIB}{Departament de F\'isica, Universitat de les Illes Balears, IAC3 -- IEEC, Crta. Valldemossa km 7.5, E-07122 Palma, Spain}
\newcommand{\ICE}{Institut de Ci\`encies de l'Espai (ICE, CSIC), Campus UAB, Carrer de Can Magrans s/n, 08193 Cerdanyola del Vall\`es, Spain}

\newcommand{\WarsawCopernicus}{Nicolaus Copernicus Astronomical Center, Polish Academy of Sciences, Bartycka 18, 00716 Warszawa, Poland}
\newcommand{\WarsawObservatory}{Astronomical Observatory, University of Warsaw, Al. Ujazdowskie 4, 00478 Warszawa, Poland}
\newcommand{\MPIAstro}{Max Planck Institute for Astrophysics, Karl-Schwarzschild-Straße 1, 85748 Garching b. München, Germany}


%% file: events.tex
\ExplSyntaxOn
\prop_new:N \g_aliases_prop

\prop_put:Nnn \g_aliases_prop {GW230601} {\event{GW230601\_224134}}
\prop_put:Nnn \g_aliases_prop {GW230605} {\event{GW230605\_065343}}
\prop_put:Nnn \g_aliases_prop {GW230606} {\event{GW230606\_004305}}
\prop_put:Nnn \g_aliases_prop {GW230608} {\event{GW230608\_205047}}
\prop_put:Nnn \g_aliases_prop {GW230609} {\event{GW230609\_064958}}
\prop_put:Nnn \g_aliases_prop {GW230624} {\event{GW230624\_113103}}
\prop_put:Nnn \g_aliases_prop {GW230627} {\event{GW230627\_015337}}
\prop_put:Nnn \g_aliases_prop {GW230628} {\event{GW230628\_231200}}
\prop_put:Nnn \g_aliases_prop {GW230630_12} {\event{GW230630\_125806}}
\prop_put:Nnn \g_aliases_prop {GW230630_23} {\event{GW230630\_234532}}
\prop_put:Nnn \g_aliases_prop {GW230702} {\event{GW230702\_185453}}
\prop_put:Nnn \g_aliases_prop {GW230704_02} {\event{GW230704\_021211}}
\prop_put:Nnn \g_aliases_prop {GW230704_21} {\event{GW230704\_212616}}
\prop_put:Nnn \g_aliases_prop {GW230706} {\event{GW230706\_104333}}
\prop_put:Nnn \g_aliases_prop {GW230707} {\event{GW230707\_124047}}
\prop_put:Nnn \g_aliases_prop {GW230708_05} {\event{GW230708\_053705}}
\prop_put:Nnn \g_aliases_prop {GW230708_23} {\event{GW230708\_230935}}
\prop_put:Nnn \g_aliases_prop {GW230709} {\event{GW230709\_122727}}
\prop_put:Nnn \g_aliases_prop {GW230712} {\event{GW230712\_090405}}
\prop_put:Nnn \g_aliases_prop {GW230723} {\event{GW230723\_101834}}
\prop_put:Nnn \g_aliases_prop {GW230726} {\event{GW230726\_002940}}
\prop_put:Nnn \g_aliases_prop {GW230729} {\event{GW230729\_082317}}
\prop_put:Nnn \g_aliases_prop {GW230731} {\event{GW230731\_215307}}
\prop_put:Nnn \g_aliases_prop {GW230803} {\event{GW230803\_033412}}
\prop_put:Nnn \g_aliases_prop {GW230805} {\event{GW230805\_034249}}
\prop_put:Nnn \g_aliases_prop {GW230806} {\event{GW230806\_204041}}
\prop_put:Nnn \g_aliases_prop {GW230811} {\event{GW230811\_032116}}
\prop_put:Nnn \g_aliases_prop {GW230814_06} {\event{GW230814\_061920}}
\prop_put:Nnn \g_aliases_prop {GW230814_23} {\event{GW230814\_230901}}
\prop_put:Nnn \g_aliases_prop {GW230819} {\event{GW230819\_171910}}
\prop_put:Nnn \g_aliases_prop {GW230820} {\event{GW230820\_212515}}
\prop_put:Nnn \g_aliases_prop {GW230824} {\event{GW230824\_033047}}
\prop_put:Nnn \g_aliases_prop {GW230825} {\event{GW230825\_041334}}
\prop_put:Nnn \g_aliases_prop {GW230831} {\event{GW230831\_015414}}
\prop_put:Nnn \g_aliases_prop {GW230904} {\event{GW230904\_051013}}
\prop_put:Nnn \g_aliases_prop {GW230911} {\event{GW230911\_195324}}
\prop_put:Nnn \g_aliases_prop {GW230914} {\event{GW230914\_111401}}
\prop_put:Nnn \g_aliases_prop {GW230919} {\event{GW230919\_215712}}
\prop_put:Nnn \g_aliases_prop {GW230920} {\event{GW230920\_071124}}
\prop_put:Nnn \g_aliases_prop {GW230922_02} {\event{GW230922\_020344}}
\prop_put:Nnn \g_aliases_prop {GW230922_04} {\event{GW230922\_040658}}
\prop_put:Nnn \g_aliases_prop {GW230924} {\event{GW230924\_124453}}
\prop_put:Nnn \g_aliases_prop {GW230927_04} {\event{GW230927\_043729}}
\prop_put:Nnn \g_aliases_prop {GW230927_15} {\event{GW230927\_153832}}
\prop_put:Nnn \g_aliases_prop {GW230928} {\event{GW230928\_215827}}
\prop_put:Nnn \g_aliases_prop {GW230930} {\event{GW230930\_110730}}
\prop_put:Nnn \g_aliases_prop {GW231001} {\event{GW231001\_140220}}
\prop_put:Nnn \g_aliases_prop {GW231004} {\event{GW231004\_232346}}
\prop_put:Nnn \g_aliases_prop {GW231005_02} {\event{GW231005\_021030}}
\prop_put:Nnn \g_aliases_prop {GW231005_09} {\event{GW231005\_091549}}
\prop_put:Nnn \g_aliases_prop {GW231008} {\event{GW231008\_142521}}
\prop_put:Nnn \g_aliases_prop {GW231014} {\event{GW231014\_040532}}
\prop_put:Nnn \g_aliases_prop {GW231018} {\event{GW231018\_233037}}
\prop_put:Nnn \g_aliases_prop {GW231020} {\event{GW231020\_142947}}
\prop_put:Nnn \g_aliases_prop {GW231028} {\event{GW231028\_153006}}
\prop_put:Nnn \g_aliases_prop {GW231029} {\event{GW231029\_111508}}
\prop_put:Nnn \g_aliases_prop {GW231102} {\event{GW231102\_071736}}
\prop_put:Nnn \g_aliases_prop {GW231104} {\event{GW231104\_133418}}
\prop_put:Nnn \g_aliases_prop {GW231108} {\event{GW231108\_125142}}
\prop_put:Nnn \g_aliases_prop {GW231110} {\event{GW231110\_040320}}
\prop_put:Nnn \g_aliases_prop {GW231113_12} {\event{GW231113\_122623}}
\prop_put:Nnn \g_aliases_prop {GW231113_20} {\event{GW231113\_200417}}
\prop_put:Nnn \g_aliases_prop {GW231114} {\event{GW231114\_043211}}
\prop_put:Nnn \g_aliases_prop {GW231118_00} {\event{GW231118\_005626}}
\prop_put:Nnn \g_aliases_prop {GW231118_07} {\event{GW231118\_071402}}
\prop_put:Nnn \g_aliases_prop {GW231118_09} {\event{GW231118\_090602}}
\prop_put:Nnn \g_aliases_prop {GW231119} {\event{GW231119\_075248}}
\prop_put:Nnn \g_aliases_prop {GW231123} {\event{GW231123\_135430}}
\prop_put:Nnn \g_aliases_prop {GW231127} {\event{GW231127\_165300}}
\prop_put:Nnn \g_aliases_prop {GW231129} {\event{GW231129\_081745}}
\prop_put:Nnn \g_aliases_prop {GW231206_2331} {\event{GW231206\_233134}}
\prop_put:Nnn \g_aliases_prop {GW231206_2339} {\event{GW231206\_233901}}
\prop_put:Nnn \g_aliases_prop {GW231213} {\event{GW231213\_111417}}
\prop_put:Nnn \g_aliases_prop {GW231221} {\event{GW231221\_135041}}
\prop_put:Nnn \g_aliases_prop {GW231223_03} {\event{GW231223\_032836}}
\prop_put:Nnn \g_aliases_prop {GW231223_07} {\event{GW231223\_075055}}
\prop_put:Nnn \g_aliases_prop {GW231223_20} {\event{GW231223\_202619}}
\prop_put:Nnn \g_aliases_prop {GW231224} {\event{GW231224\_024321}}
\prop_put:Nnn \g_aliases_prop {GW231226} {\event{GW231226\_101520}}
\prop_put:Nnn \g_aliases_prop {GW231230} {\event{GW231230\_170116}}
\prop_put:Nnn \g_aliases_prop {GW231231} {\event{GW231231\_154016}}
\prop_put:Nnn \g_aliases_prop {GW240104} {\event{GW240104\_164932}}
\prop_put:Nnn \g_aliases_prop {GW240107} {\event{GW240107\_013215}}
\prop_put:Nnn \g_aliases_prop {GW240109} {\event{GW240109\_050431}}
\prop_put:Nnn \g_aliases_prop {GW241011} {\event{GW241011\_233834}}
\prop_put:Nnn \g_aliases_prop {GW241110} {\event{GW241110\_124123}}
\prop_put:Nnn \g_aliases_prop {GW250114} {\event{GW250114\_082203}}

\ExplSyntaxOff